\documentclass[useAMS,usenatbib]{mn2e}
\usepackage{graphics}
\usepackage{subfigure}
\usepackage{graphicx}
\usepackage{amssymb}
\usepackage{hyperref}
\title[
Black hole spin and radio loudness 
]{
Black hole spin and radio loudness in a $\Lambda$CDM universe
} 
\author[Claudia del P. Lagos, Nelson D. 
Padilla, Sof\'ia A. Cora]{Claudia del P. Lagos$^{1}$, Nelson D.
Padilla$^{1}$, Sof\'ia A. Cora$^{2,3}$\\
$^{1}$Departamento Astronom\'ia y Astrof\'isica, Pontificia Universidad
Cat\'olica de Chile, Av. Vicu\~na Mackenna 4860, Stgo., Chile\\
$^{2}$ Facultad de Ciencias Astron\'omicas y Geof\'isicas de la Universidad Nacional
de La Plata, and Instituto de Astrof\'isica de La Plata \\(CCT La Plata, CONICET, UNLP), Observatorio Astron\'omico, Paseo del Bosque S/N, 1900 La Plata, Argentina\\
$^{3}$Consejo Nacional de Investigaciones Cient\'{\i}ficas y T\'ecnicas,
Rivadavia 1917, Buenos Aires, Argentina\\
}
\begin{document}

\date{Accepted ???. Received ???; in original form ???}

\pagerange{\pageref{firstpage}--\pageref{lastpage}} \pubyear{2008}

\maketitle

\label{firstpage}

\begin{abstract}
We use
a combination of a cosmological {\em N}-body simulation of the
concordance $\Lambda$
cold dark matter ($\Lambda$CDM) paradigm and a semi-analytic model of
galaxy formation to investigate the spin development of central
supermassive black holes (BHs) and its 
relation to the BH host galaxy properties.
In order to compute BH spins, we use the $\alpha$ model
of Shakura \& Sunyaev and consider the King et al.
warped disc alignment criterion. 
The orientation of the accretion disc is 
inferred from the angular momentum of
the source of accreted material, which
bears a close relationship to the large-scale structure in the simulation. 
We find that the final BH spin depends almost exclusively on the accretion
history and only weakly on the warped disc alignment.
The main mechanisms of BH spin-up are found to be gas
cooling processes and disc instabilities, a result that is only
partially compatible with Monte-Carlo models where the main spin-up mechanisms
are major mergers and disc instabilities; 
the latter results are reproduced
when implementing
randomly oriented accretion discs in our model.
Regarding the BH population, we find that more massive BHs, which are 
hosted by massive
ellipticals, have higher spin values than less-massive BHs, hosted by
spiral galaxies.
We analyse whether gas accretion rates and BH spins can be used as 
tracers of the radio loudness of active galactic nuclei (AGN). We find that 
the current observational indications of an increasing trend of radio-loud
AGN fractions with stellar and BH mass can be easily obtained
when placing lower limits on the BH spin, with a minimum influence 
from limits on the accretion rates; a model with random accretion disc
orientations is unable to reproduce this trend. Our results favour
a scenario where the BH spin is a key parameter to separate
the radio-loud and radio-quiet galaxy populations.
\end{abstract}

\begin{keywords}
galaxies: evolution - galaxies: active - physical processes: accretion discs
\end{keywords}

\section{Introduction}\label{Introsec}

Galaxies with Active Galactic Nuclei (AGN) have become the centre of attention
in extragalactic studies due to their star role in the galaxy formation scenario
favoured today.  They are a key ingredient in explaining 
several observational statistics on
galaxy star formation rates, luminosities and colours
(see for instance,
\citealt{Bower06}, \citealt{Croton06}, \citealt{Cattaneo06},
\citealt{Sijacki07}, \citealt{Marulli08}, \citealt{Lagos08}, \citealt{Somerville08}).
However, a number of phenomena related to AGN are still unclear, including
the origin of the AGN dichotomy into
``radio loud'' (RL) and ``radio quiet'' (RQ) types. 
RL AGN are characterised by the presence
of jets of relativistic plasma and/or associated large-scale radio
lobes. 
Relativistic jets
emitting synchrotron radiation in the radio band
are generated by the collimation of outflows from the innermost
regions of accretion discs around black holes (BHs), carrying away
a large fraction of the available accretion power. However,
the jet production mechanism
is not fully understood.
The double-lobed morphology \citet{Fanaroff74} Class II
(FR II) sources are 
the most common type observed in broad line emitting RL
quasars.
On the other hand, the lower luminosity Fanaroff Class I objects (FRI) 
have very weak
radio emitting ejecta and  
weak or no emission lines.
The correlation between radio luminosity and line luminosity in $[\rm OIII] \,\lambda 5007$  
is similar for both RL and RQ AGN
\citep{Xu99}.

Different criteria for the definition of radio loudness have been used, 
the most common being the
ratio between radio-to-optical luminosity, ${\cal R}$ \citep{Kellermann89},
and the radio power \citep{Miller90}.
A third criterion based on the classical radio morphology
has also been considered \citep{Sulentic03}.
From the combination of the two latter criteria,
\citet{Zamfir08} find that 
the boundary between RL and RQ AGN is given by 
${\rm log}\, L_{1.4 \, {\rm
GHz}}=31.6\,{\rm erg}\,{\rm s^{-1}}\,{\rm Hz^{-1}}$, 
which corresponds to the luminosity of 
the weakest FRII sources.
Using this luminosity threshold, they obtain a fraction 
of $\sim 5 - 9$ per cent of RL AGN relative to the total population
of AGN. 
The presence of a physical bimodality in the properties of AGN 
is still a matter of debate, 
and it should be borne in mind that observational selections
might actually
induce this dichotomy
(e.g. \citealt{Cirasuolo03}, \citealt{White07}).  

The host galaxies of both RL and RQ QSOs with nuclear luminosities $M_V < -24$ 
are massive ellipticals
(\citealt{Dunlop03}, \citealt{Floyd04});
disc-dominated host galaxies
become increasingly rare as the nuclear power increases.
Sikora, Stawarz \& Lasota (2007)
demonstrate that radio-selected AGN hosted by
giant ellipticals can be up to a thousand times brighter in radio frequencies than AGN hosted by 
disc galaxies, thus generating two sequences in the 
radio versus optical luminosity plane.
These RL and RQ sequences remain present
when considering the dependence of the radio loudness ${\cal R}$ with the 
Eddington ratio, where  
the latter is defined as the
bolometric luminosity (estimated from optical data) in units
of the Eddington luminosity. 
The two sequences show an increase of
radio loudness with decreasing Eddington ratio.  This bimodality breaks
down at high accretion rates.
This behaviour was
originally noticed by \citet{Ho02}, although only a single sequence was
visible as a consequence of the reduced sample of AGN considered.

The link between jet production
and gas accretion rate onto the BH has been widely investigated.
The inner part of the accretion flow is believed to be responsible for
the observed hard X-ray emission produced by Compton scattering
or Bremsstrahlung radiation.
Thus, observations of the compact emission in the X-ray and radio bands
help understand 
the disc-jet connection in both stellar and 
supermassive BHs 
(\citealt{Ulvestad01}, \citealt{Merloni03}, \citealt{Falcke04}, 
\citealt{Kording06}),
under the assumption that the jet formation process
is qualitatively similar on both scales.
Observations of black hole binary systems during the 
low/hard X-ray state reveal a tight correlation
between the radio and X-ray luminosity \citep{Gallo03}, supporting
the dependence of the radio power on the accretion rate.
This ``acretion paradigm'' postulates that
radio loudness is
entirely related to the states of accretion discs.  However, it 
cannot explain the 
two parallel sequences AGN occupy
in the plane defined by radio loudness and Eddington ratio, as
found by \citet{Sikora07}.  Hence, it is possible that additional 
BH properties 
could be playing an important role in
driving the 
production of powerful relativistic radio jets.

The RL and RQ sequences found by
\citet{Sikora07} are characterised by AGN containing
BHs with masses $\gtrsim 10^{8}\,M_{\odot}$
and $ \la 3\times10^{7}\,M_{\odot}$,
respectively.  Nonetheless, the BH mass is not capable of separating
completely these two sequences, since
objects powered by equally massive BHs
can differ in radio luminosity by $\sim 4$ orders of magnitude
(\citealt{Ho02}, \citealt{Dunlop03}, \citealt{Sikora07}).
\citet{Blandford90} introduces a new parameter
suggesting that the efficiency of jet
production is determined by the dimensionless BH spin,
according to the assumption that relativistic jets
are powered by rotating BHs through the Blandford-Znajek
mechanism.
In an attempt to explain this radio power dichotomy at
fixed accretion rates, \citet{Sikora07} propose a revised version of
the ``spin paradigm'' which
suggests that giant elliptical galaxies host, on average, BHs with 
larger spins
than those hosted by spiral galaxies.
These new insights inferred from observational evidences motivated
the study of the influence from 
BH mergers and small accretion episodes on the evolution of BH spins
and their relation to the morphology of the host galaxy
(e.g. \citealt{Volonteri07}; \citealt{Nemmen07}; \citealt{Berti08}).
These theoretical works indicate that it is 
feasible
that the observed morphology-related bimodality
of AGN radio loudness arises from the dichotomy shown by the
distribution of BH spins.

In this paper, we investigate the latter possibility
by using the SAG { (acronym for `Semi-Analytic Galaxies')}
semi-analytic model of galaxy formation 
by Lagos, Cora \& Padilla (2008, hereafter LCP08),
combined with the outputs of a cosmological
simulation based on the
$\Lambda$CDM cosmology. 
The advantage of this approach relies in that we are able to use 
realistic BH growth histories directly 
linked to the evolution of the star formation rate
of their host galaxies. The SAG model also allows us to evaluate
the influence of the different 
mechanisms that contribute to the accretion of material onto the BHs, which
drive both
the BH mass growth and the BH spin development.

This work is organised as follows. 
In Section~\ref{modeltot}, we briefly summarise
the mechanisms that contribute to the BH growth included in
SAG, and describe in detail the 
implementation of the BH spin model. 
In Section~\ref{nature}, we discuss the connection between
the BH spin distributions and the   
characteristics of the BH host galaxies.
Section~\ref{growth} shows the results on the final BH spin
and the way in which it is acquired. 
In Section~\ref{radioloud}, we analyse 
the nature of AGN radio loudness and compare our results with 
available observational data.
Finally, 
the main conclusions obtained in this work are summarised 
in Section~\ref{conclusion}.

\section{Computing the black hole spin}\label{modeltot}

We study the
relationship between the BH spin and the properties of the BH host galaxy
using a combination of a cosmological {\em N}-Body
simulation of the concordance $\Lambda$CDM universe
and the SAG semi-analytic model of galaxy formation
described in LCP08.
This model considers radiative cooling of hot gas, star formation, 
galaxy mergers, disc instabilities, 
feedback from supernova explosions, 
BH growth and AGN feedback produced during accretion onto BHs driven
by gas cooling processes. 

In the following subsections, we briefly describe the $N$-Body cosmological
simulation and the BH growth model used in SAG, relevant for
the present study. We then explain in detail the implementation of the
model to compute the BH spin.

\subsection{$\Lambda$CDM Cosmological simulation} \label{SimuLambda}

The cosmological simulation is the same as the one used by LCP08, 
based on the concordance
$\Lambda$CDM cosmology. It considers a periodic box of $60 \, h^{-1}\,{\rm Mpc}$
containing
$16,777,216$ dark matter particles with mass $1.001 \times
10^{9}\,h^{-1}\,{\rm M}_{\odot}$. It has more than $54,000$ dark matter haloes with masses
of up to $5.36 \times 10^{14}\,h^{-1}\,{\rm M}_{\odot}$.
The simulation parameters are consistent with the results of
WMAP data
\citep{spergel03}, with  $\Omega_{\rm m}=\Omega_{\rm DM}+\Omega_{\rm
baryons}=0.28$ (with a baryon fraction of $0.16$), $\Omega_{\Lambda}=0.72$ and
$\sigma_{8}=0.9$. The Hubble constant is set to
$H_{0}=100 \, h\, {\rm Mpc}^{-1}$, with $h=0.72$, and the gravitational softening
length is  $\epsilon=3.0 \,h^{-1}\,{\rm Kpc}$.
The simulation starts from a redshift
$z=48$ and has been run using the public version of the {\small GADGET-2} code
 \citep{Springel05}.

\subsection{Black hole growth}\label{BHgrowth}

There are { three} distinct modes of BH growth in SAG. The first one, 
referred
to as the ``QSO mode'',
is associated to starbursts in galaxies. 
These events can be driven by galaxy minor or
major mergers and the collapse of unstable discs. The second one,
referred to as the ``radio mode'', is associated to gas cooling processes
that produce star formation in a more quiescent way. { The third one is 
BH mergers which occur shortly after the merger of the parent galaxies.}
In our model, AGN feedback 
occurs only during the latter 
mode which produces low accretion rates, in agreement with observations
(e.g. \citealt{Ho02}; \citealt{Donahue05}; \citealt{Sikora07}).
The detailed
assumptions for each mechanism are given in LCP08. 
Here, we briefly describe the main formulae involved.

During galaxy mergers, the BH growth depends on the ratio
between total galaxy mass and the amount of available cold gas {
(\citealt{Croton06}, LCP08)} as 

\begin{equation}
        \dot{M}_{\rm BH}= \frac{f_{\rm BH}}{\Delta t} \frac{M^{\rm sat}}{M^{\rm central}}
\times
\frac{M_{\rm ColdGas}}{1+(200 \,{\rm km}\, {\rm s}^{-1}/V_{\rm vir})^{2}},
\label{massqsomode}
\end{equation}

\noindent where
$M_{\rm ColdGas}$
is the cold gas mass of the system, consisting of the central and satellite
galaxies, with masses (including stellar and cold gas mass)
$M^{\rm sat}$ and $M^{\rm central}$,
respectively;
$\Delta t$ is the timescale between consecutive SAG timesteps.
 Black holes are assumed to grow with an efficiency $f_{\rm BH}=0.015$, {
set to fit the $M_{\rm BH}$-$M_{\rm Bulge}$ relation}. 
For disc
instabilities, the ratio $M^{\rm sat}/M^{\rm central}$ is replaced by unity,
since this process depends on
the properties of a single galaxy.
It is important to note that there are no BH seeds in our model and,
therefore, the birth of central supermassive BHs is triggered
when galaxies undergo their first starburst at some point in their evolution.

The accretion rate during the ``radio mode'' is given by
\begin{equation}
        \dot{M}_{\rm BH}= \kappa_{\rm AGN} \frac{M_{\rm BH}}{10^{8}
M_{\odot}} \times \frac{f_{\rm hot}}{0.1} \times \left(\frac{V_{\rm vir}}{200\,
{\rm km}\,
{\rm s}^{-1}}\right)^{3} \label{massradio},
\end{equation}
\noindent where 
$M_{\rm BH}$ is the BH mass, and $f_{\rm hot}$ is the
fraction of the total halo mass in the form of hot gas, $f_{\rm hot}=m_{\rm
HotGas}/M_{\rm vir}$, $M_{\rm vir}$ being the virial mass of the host halo
which has a virial velocity $V_{\rm vir}$;
$\kappa_{\rm AGN}$ is a free parameter
{ set to fit the galaxy
and QSO luminosity functions} with a value $2.5\times10^{-4} \,{\rm M}_{\odot}\, {\rm yr}^{-1}$. 

Irrespective of the accretion mode, the BH produces a luminosity
that corresponds to the total release of energy from the disc
formed by gas accretion.  This
luminosity is given by $L_{\rm BH}=\eta \, \dot{M}_{\rm BH} c^{2}
\label{LBH}$, where  $c$ is the speed of light, and $\eta=0.1$ is the standard
efficiency of energy production that occurs
in the vicinity of the event horizon (\citealt{Shakura73}, SS73 hereafter).

\subsection{Black hole spin}\label{model}

Our implementation of a BH spin model in SAG follows 
the same methodology adopted
for all the galaxy properties considered in the semi-analytic model, 
which consists in solving 
a set of linearised differential equations.
The BH spin in
each galaxy is followed through the evolutionary history of the
simulated universe, allowing us to 
analyse
the way in which the BH spin is acquired for 
every host galaxy and to study directly the role and
influence of the hierarchical
build-up of structure in the development of this 
BH property.
During galaxy mergers, the two central supermassive BHs are also
merged; we simply
consider that
the mass of the final BH is the sum of the merging BH masses; { the
resulting spin is
calculated using the semi-analytic fitting formulae presented by
\citet{Rezzolla08} to fit 
a complete compilation of numerical results on BH mergers.}
The other processes involved in the BH growth
feed baryons through the formation
of accretion discs. The treatment of the angular momentum transfer
from the accretion disc to the BH requires the following analysis. 

The dimensionless BH spin is defined as \^a $\equiv J_{\rm
BH}/ J_{\rm MAX}=c
J_{\rm BH}/G M_{\rm BH}^{2}$, where $J_{\rm BH}$ is the angular momentum of the
BH.
In general, $J$ refers to the absolute value of the angular
momentum, $\vert \vec{J} \vert$.
We consider the presence of a warped disc characterised by
the angular momentum of the accretion disc at the warp radius,
$J_{\rm d}=J_{\rm d}(R_{w})$ \citep{Volonteri07}.
In this case, the amount $J_{\rm d}/ J_{\rm
BH}$
can be used to define the 
alignment
between the spinning BH and the accretion disc
(\citealt{King05}, hereafter K05).
Angular momentum vectors are co-aligned
when ${\rm cos}(\phi)>-0.5 J_{\rm d}/ J_{\rm BH}$, being
$\phi$
the angle between the disc angular momentum and the BH spin; otherwise,
the accretion disc is counter-aligned with the BH spin. 
We use this
criterion to evaluate the BH spin-up or down in the presence of a warped disc.

SAG does not assume an initial mass for BHs, or BH seeds; instead,
the first starburst in a galaxy,
that occurs during galaxy mergers or disc instabilities, 
creates the central BH. 
These processes would therefore be
in line with an origin of super-massive BHs as
remnants of the first events
of stellar explosions
early in the history of the Universe 
(e.g. \citealt{Bernadetta08}, \citealt{Omukai08},\citealt{Shapiro04}, \citealt{Elmegreen08}). 
Given this possible stellar origin of the BHs, it is reasonable to expect
their initial spin to be non-zero. However, for the sake of
completeness we also study the effect of initially spinless
BHs, which could be formed by rapid collapse of infalling
gas (see for instance \citealt{Begelman06}). In
this latter case, BHs acquire spin for the first time
during their first accretion episode or merger with
other black hole. In the case of an
accretion event, the resulting spin is given by the formalism developed by
\citet{Bardeen70}. When two spinless BHs merge, we use the analytic formula given
by \citet{Berti08}; when at least one
merging BH has a non-zero spin,
the resulting
spin is calculated using the semi-analytic fitting formulae
presented by \citet{Rezzolla08} to fit a complete
compilation of numerical results on BH mergers.
 
{ In the case of BHs with non-zero spin, we calculate 
$J_{\rm d}$ taking into account the mass of the accretion disc, $M_{\rm d}$,} and the remaining
properties related to the BH, $M_{\rm BH}$, $\dot{M}_{\rm BH}$ and
$L_{\rm BH}$; all these quantities are provided by the SAG model.
The ratio between the angular momentum of the accretion disc and the BH spin
is expressed as 
\begin{equation}
\frac{J_{\rm d}}{J_{\rm BH}}=\frac{M_{\rm d}}{\hat{\rm a} M_{\rm BH}}
\left(\frac{R_{w}}{R_{s}}\right)^{1/2},
\label{e1}
\end{equation}
\noindent where $M_{\rm d}=\dot{M_{\rm BH}} \Delta t$. 
The accretion rate $\dot{M_{\rm BH}}$ comes either from
equation ~\ref{massqsomode} or ~\ref{massradio}. 
Accretion discs are completely consumed in the time interval $\Delta t$.
The ratio $R_{w}/R_{s}$ is the warp radius expressed in terms
of the Schwarzschild radius, $R_s$; in a SS73 middle-region disc, this ratio can be
written as
\begin{equation}
\frac{R_{w}}{R_{s}}=3.6 \times 10^{3} \hat{\rm a}^{5/8} \left(\frac{M_{\rm BH}}{10^{8}
M_{\odot}}\right)^{1/8} f_{\rm Edd}^{-1/4}
\left(\frac{\nu_{2}}{\nu_{1}}\right)^{-5/8} \alpha^{-1/2}
\label{e2}
\end{equation}
\noindent (e.g. K05, \citealt{Volonteri05}), where
$f_{\rm Edd} \equiv L_{\rm BH}/L_{\rm Edd}$, being
$L_{\rm Edd}$ the Eddington luminosity, $\nu_{2}$ the
warp propagation viscosity, and $\nu_{1}$ the accretion driving
viscosity ($\nu_{2}$ can be different from $\nu_{1}$).
The parameter $\alpha$, introduced by
SS73, takes into account the efficiency of the mechanism of
angular momentum transport.
A thin accretion disc is characterised by $H/R < \alpha \ll 1$, where
$H$ and $R$ are the disc thickness and radius, respectively.
In this case, it can be shown that $\nu_{2}/\nu_{1} \approx \alpha^{2}$
\citep{Papaloizou83}.
Nonetheless, at
high accretion rates (i.e. hot/thick accretion discs) there
are indications that $\alpha \gtrsim 0.1$ (e.g. \citealt{Dubus01}).
Based on these theoretical and observational
indications, we adopt simple criteria
to distinguish between thin and thick accretion discs and, therefore,
to choose an appropriate value for $\alpha$:
\begin{itemize}
\item {\it Thick accretion disc}: if the accretion occurs at super-Eddington
rates, that is
$L_{\rm BH}/L_{\rm Edd} > 1$, 
the accretion proceeds via a radiation pressure supported thick disc with
$H \, \sim R$ (e.g. \citealt{Volonteri05}; \citealt{Begelman06}), 
which implies
$\nu_{1}\approx \nu_{2}$ (e.g. \citealt{Kumar85})
 and $\alpha \gtrsim 0.1$
(e.g. \citealt{Dubus01}). 
\item {\it Thin accretion disc}: 
if the accretion occurs in the regime of
$L_{\rm BH}/L_{\rm Edd} < 1$, it is assumed that
$H/R \sim \alpha \ll 1$.
\end{itemize}

In order to test
the reliability of our results, 
we adopt different values of the parameter $\alpha$,
ranging from $\sim 0.1$ to $\sim 0.5$, for the thick disc, 
and from $\sim 10^{-4}$ to $\sim 0.05$, for the thin disc.
It is important to remark that the distinction 
between thick and thin discs has
a direct consequence 
on
the amount of angular momentum per unit mass transferred  
from the accretion disc to the BH, 
for the same disc
mass. This arises from the dependence of equation \ref{e2} on 
the parameter $\alpha$, and the 
viscosities $\nu_{1}$ 
and $\nu_{2}$.
This is a simplified approach to more sophisticated models which 
emphasize the importance of the disc thickness 
(e.g. \citealt{Hawley06}).

There are a number of works
that study in detail the possibility of alignments between galactic-scale
structures and the central BH-disc system.  \citet{Bogdanovic07} find 
indications of co-alignment between the BH spin and
large-scale gas flows coming from galaxy mergers. 
Also, \citet{Lindner07} show that the jet direction is set by
the BH angular
momentum.  
These results indicate a possible co-alignment 
between the jet direction and
the angular momentum of the accretion disc of the host galaxy,
also supported by observations of high redshift radio galaxies
(\citealt{Chambers87}; \citealt{Lacy99}; \citealt{Inskip05}). 
Moreover,
\citet{Berti08} show that high spin values are efficiently achieved by a model
with non-chaotic accretion episodes, that could
explain the very large spin $\hat{a} \approx 0.99$ of the case of MCG-06-30-15
\citep{Brenneman06}.
{These arguments make it
relevant to consider the possibility that the direction
of the angular momentum of the accretion disc is related 
to the physical process that triggers the
gas accretion.}

{We test this possibility using 
the information provided by
our model as follows.
When the accretion is driven by disc instabilities,
we assume that the
direction of the accretion disc
can be inferred from the angular momentum}
$\vec{L}_{\rm Gal,disc}$ of the galaxy disc,  

\begin{equation}
\vec{J_{\rm d}} = J_{\rm d} \cdot \hat{L}_{\rm Gal,disc}. 
\label{J1}
\end{equation}

\noindent {On the other hand, if the accretion is driven by 
gas cooling, we assume that the direction will be that of 
the angular momentum of the host halo,}

\begin{equation}
\vec{J_{\rm d}} = J_{\rm d} \cdot \hat{L}_{\rm Halo}. 
\label{J1}
\end{equation} 

\noindent 
If the
gas accretion is driven by galaxy mergers, the direction is
taken from the gas mass-weighted 
disc angular momentum contributed by each galaxy, 

\begin{eqnarray}
\vec{J_{\rm d}} = J_{\rm d} \cdot \left(\hat{L}_{\rm Gal,disc}^{\rm central} 
\times \frac{M_{\rm ColdGas}^{\rm central}}{M_{\rm ColdGas}} + \hat{L}_{\rm Gal,disc}^{\rm sat} \times \frac{M_{\rm ColdGas}^{\rm sat}}{M_{\rm ColdGas}}\right), 
\label{J2}
\end{eqnarray}

\noindent where 
$M_{\rm ColdGas}^{\rm central}$ and
$M_{\rm ColdGas}^{\rm sat}$ 
are the cold gas mass of the
central and satellite galaxies, respectively.
Having the information on $\vec{J}_{\rm d}$, 
we determine the co-(counter-)alignment
according to the K05 criteria
to finally sum (subtract) $J_{\rm d}$ to $J_{\rm BH}$. 

{Since a $\Lambda$CDM universe is characterised
by strong co-alignments between the angular momentum of dark-matter haloes
and their surrounding structures on scales of several Mpc (\citealt{Paz08}),
there could be some level of alignment between the BH spin and the large-scale
structure.  In order to explore this,
the initial angular momentum of the gaseous disc of our model galaxies are
inferred from that of their host dark-matter haloes. 
Then, the angular momentum of model galaxies is followed and updated at
every star formation
episode, which can introduce some degree of misalignment with that
of the hosting haloes.
In the case of gas cooling processes, we consider that the cooled gas
added to the galaxy disc drags the angular momentum of the host halo
\citep{Cole00}, modifying the pre-existing angular momentum of the galaxy disc.
During mergers and disc instabilities, the contributions from the
participating gas and stellar components
are taken into account to obtain the resulting galaxy angular momenta.
}
We also consider
randomly oriented accretion events (e.g., \citealt{King07};
\citealt{Volonteri07}) in order to
study the main consequences arising from these two
possible scenarios.

We also analyse the results obtained from a model 
where the timescale for disc warping is much longer than for gas accretion.
In this case, we ignore the K05 alignment criterion
and 
the angular momentum of the disc is simply vectorially
added to the BH spin.
In all cases, the final BH spin direction will be that of the
total angular momentum of the system (BH and accretion disc) which is
assumed to be conserved. This assumption is not fully accurate since the torque 
exerted
by the accretion disc acting over the total angular momentum of the system may 
be important (e.g. \citealt{Volonteri05}). 

Since we are not considering angular momentum losses 
by torques,
our estimates of the values of
BH spins are
upper limits. 
In case that the spin takes unphysical values (\^a $> 1$),
we simply set them to \^a $= 1$. However, when the resulting 
spin values
are more than half a decade larger than unity, we
also reduce the gas accretion onto the BH consistently with this new upper
limit.  
We find that adopting
values of $\alpha=0.5$
for thick discs and $\alpha=0.05$ for thin discs, 
only $15$ per cent
of all accretion events in our model produces
values of \^a$\,>1$ (only $\approx 0.1$ per cent yields 
\^a$\,>3$).
Lower values of $\alpha$ produce higher
percentages of unphysical spin values, 
as can be inferred from equations \ref{e1} and \ref{e2}.
Therefore, our choice of $\alpha$ produces a minimum bias on the resulting spin 
distributions (a high concentration at \^a$\,\approx1$ would spuriously hide 
information
on the high tail of the BH spin distribution).

One of the main uncertainties in our spin model lies on 
the initial value of the BH spin when considering their possible stellar origin,
for which we test two possibilities, (i) a constant initial
value (from $\hat{a}_{\rm initial}=10^{-4}$ to $0.5$\footnote{{ These initial
spin values are
consistent with results from hydrodynamic simulations (see \citealt{Shapiro04}
and references therein).}}), and (ii) 
an initial value proportional to the BH mass. As we will show 
in Section~\ref{nature}, 
the choice of the initial spin value does not affect our 
results significantly. Moreover, the general conclusions of this work 
are preserved even for the case of
spinless initial BHs. The most sensitive population to the inclusion of an initial BH
spin are low-mass BHs (i.e. $M_{\rm BH}<10^6
M_{\odot}$) since most of them have very few accretion episodes and practically no merger
events during their entire lifetimes; as a note of caution, 
we remind the reader 
that a null initial spin is not likely to occur in a BH
with a stellar origin.

In the case when a BH is assigned an initial spin (appart from a merger event)
the spin direction will depend on
the process that triggers the BH birth.
Thus, the direction is given by 
(i) the angular
momentum of the galaxy disc 
(similar to Eq.~\ref{J1}),
in the case of disc instabilities,
and (ii) the angular
momentum of the galaxy disc weighted by the gas mass of each merging galaxy
(similar to Eq.~\ref{J2}), in the case of galaxy mergers.
The final value of the spin will be
only a function of the accretion and assembly history
of the BH.

This model provides a complete 
characterisation of a supermassive BH, giving information on
both its mass and spin.
These quantities strongly depend on the mass of cold gas accreted onto the BH.
The great advantage of our model is that accretion discs are 
consistently given 
by our semianalytic model 
within a 
universe where structures grow hierarchically,
instead of being obtained from
the assumption of specific distributions of Eddington accretion ratios as
in other works (e.g. \citealt{Sikora07}).
Indeed, the 
redshift evolution of the accretion rate density onto a BH
found by LCP08 (see their Fig. 2) 
is consistent with the empirical estimations given by
\citet{Merloni08}. Moreover, the evolution of the QSO luminosity function 
obtained from SAG
is in good agreement with observational results 
(\citealt{Wolf03}; \citealt{Croom04}).
{ 
Consequently, the resulting distributions of accretion disc masses and Eddington rates are consistent
with what was adopted in previous Monte-Carlo models.}

\subsection{Accretion disc orientation: inferred or random?}\label{advantage}

An important 
improvement presented by this model resides in the 
use of the known, 
vectorial angular momenta of the components of model galaxies
to infer the
direction of the angular momentum of the accretion disc. 
This depends on the physical process that triggers the accretion event, that
is, gas cooling, disc instabilities, 
or galaxy mergers, as detailed above.
Here we describe the main advantages of this approach.

\begin{figure}
\begin{center}
\includegraphics[width=0.48\textwidth]{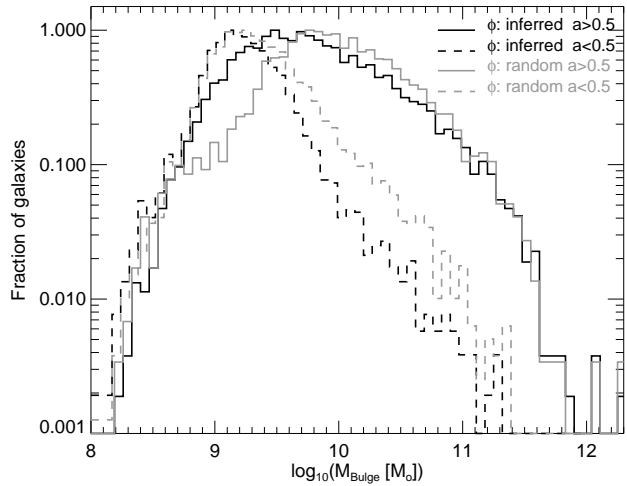}
\caption{Distributions of $z=0$ bulge masses for
different ranges of final BH spin values, \^a$\,<0.5$ (dashed lines) and \^a$\,\ge 0.5$ (solid lines).
Black lines correspond to a model where accretion
discs orientations are inferred
from galaxy discs according to 
the process that triggers the accretion event; 
grey lines denote
random 
accretion discs orientations.
Both models take into account the K05 alignment criterion and use
a constant initial black hole spin value \^a$_{\rm initial}=0.01$. 
}
\label{bulges}
\end{center}
\end{figure}
 
Fig.~\ref{bulges} shows the 
distributions of bulge masses for galaxies hosting central
BHs with low and high spin values 
at $z=0$ (dashed and solid lines, respectively),
for a model where the angular momentum of the accretion disc
depends on the process responsible for the gas accretion onto the central BH 
(Eq. \ref{J1} and \ref{J2})  
(model A, black lines), and a model that uses
randomly oriented accretion discs (model B, grey lines). 
Both models take into account the K05 alignment criterion and use
a constant initial black hole spin \^a$_{\rm initial}=0.01$. 
As can be seen, in model A, 
{the fraction of galaxies with bulges of masses $M_{\rm Bulge}\gtrsim
3 \times 10^{9} M_{\odot}$ hosting low spin BHs 
decreases faster than in model B,
while both models show similar fractions of galaxies for a given bulge mass
that host BHs with high spin values. Results from both models indicate that
massive bulges host exclusively high spin BHs.}
Taking into account 
that the most massive galaxies are ellipticals
\citep{Conselice06}, 
we find that model A, which shows
a slightly clearer connection
between massive galaxies and high BH spin values,
supports the conjecture proposed by \citet{Sikora07} that 
the radio loudness bimodality can be explained by the
morphology-related bimodality of the BH spin distribution.
Sections ~\ref{nature} and ~\ref{growth} are devoted to explore
the predictions of this model. 

\section{Connection between black hole spin distributions and host galaxy properties}\label{nature}

The main purpose of this work is to find a link between the properties of
BHs and those of their host galaxies.
In order to test the reliability of the possible relations obtained,
we evaluate their dependence on different physical assumptions 
included in our BH spin model,
such as the initial value of
the BH spin and the alignment of the accretion disc.

\begin{figure}
\begin{center}
\includegraphics[width=0.35\textwidth]{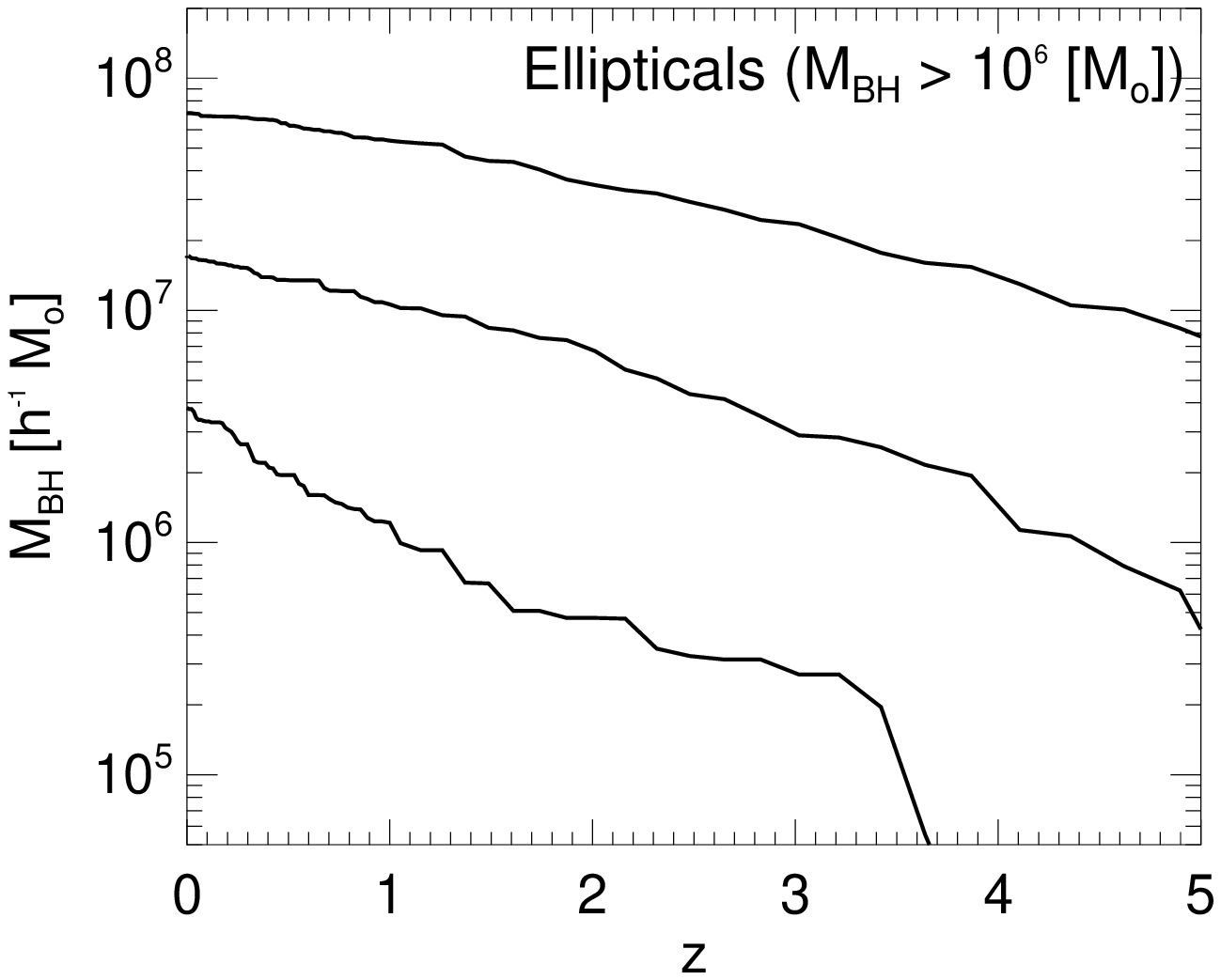}
\includegraphics[width=0.35\textwidth]{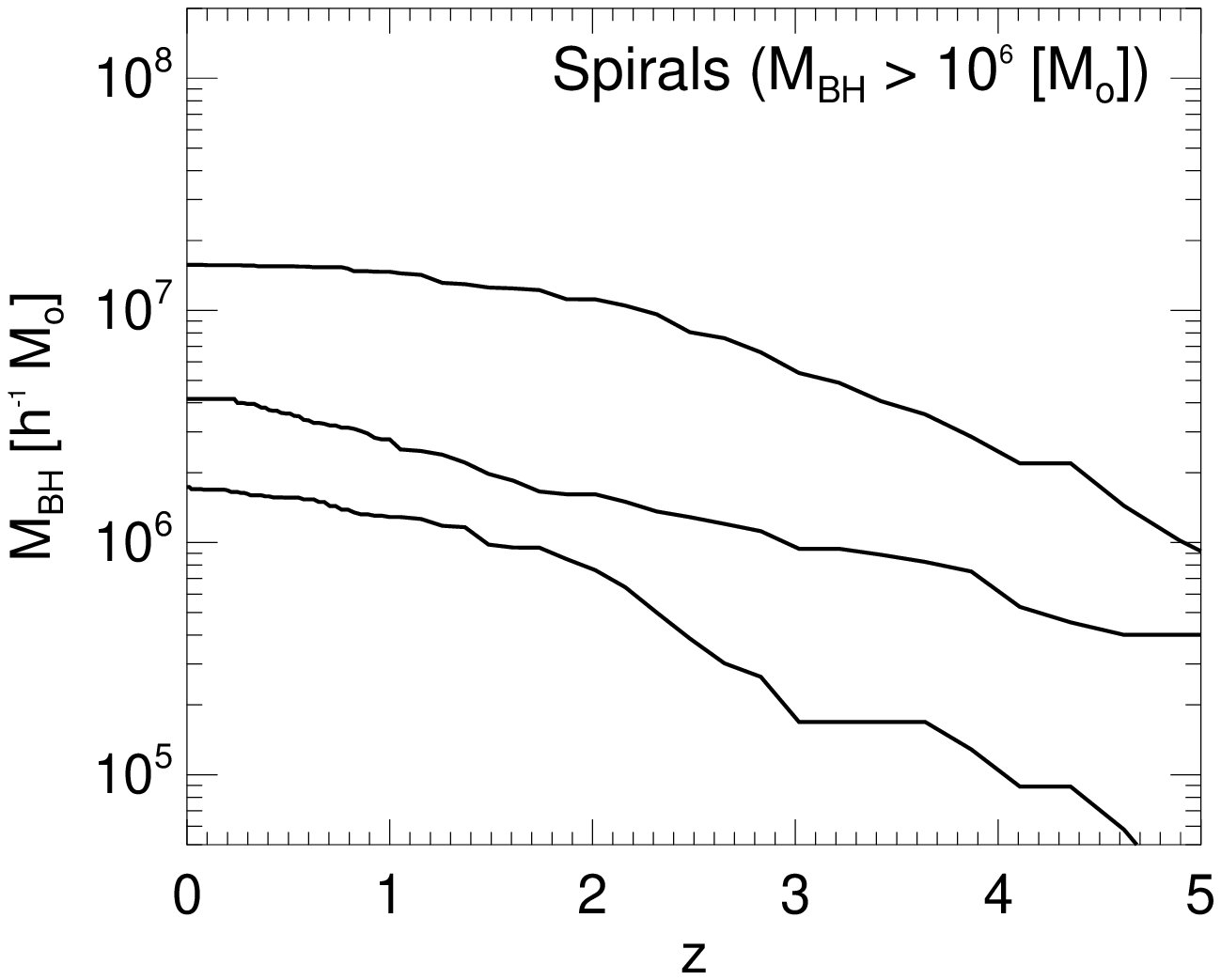}
\caption{History of average BH growth for elliptical and spiral galaxies (top and bottom
panels, respectively).
Different lines show different average final BH masses (each mass bin
contains equal number of BHs).
Elliptical galaxies are selected according to 
$M_{\rm Bulge}/M_{\rm StellarTotal} > 0.95$,
while spirals satisfy the condition
$0 < M_{\rm Bulge}/M_{\rm StellarTotal} < 0.95$.}
\label{tracks}
\end{center}
\end{figure}

Fig.~\ref{tracks} shows average growth tracks for BHs hosted by elliptical
(top) and
spiral galaxies (bottom). In both cases, we show BHs with masses
$M_{\rm BH} > 10^{6} \,M_{\odot}$, separated
in three mass bins, each containing
the same number of BHs. 
The morphological criterion to distinguish between elliptical
and spirals considers the fraction $M_{\rm Bulge}/M_{\rm StellarTotal}$
between the bulge mas and the total stellar mass of the galaxy.  { We
consider as ellipticals galaxies, those with $r=M_{\rm Bulge}/M_{\rm
StellarTotal}>r_{\rm thresh}=0.95$,
chosen in order to recover observed morphological distributions (see LCP08)}.
These tracks indicate that, at all times, ellipticals host
more massive BHs than spirals. Besides, the latter tend to stop 
the growth of their central BH at earlier times.
These results 
do not depend on the details of the BH spin model.
The differences in final mass and assembly history of the BHs can
produce different final BH spins, a possibility that we now turn to.

\begin{figure}
\begin{center}
\includegraphics[width=0.48\textwidth]{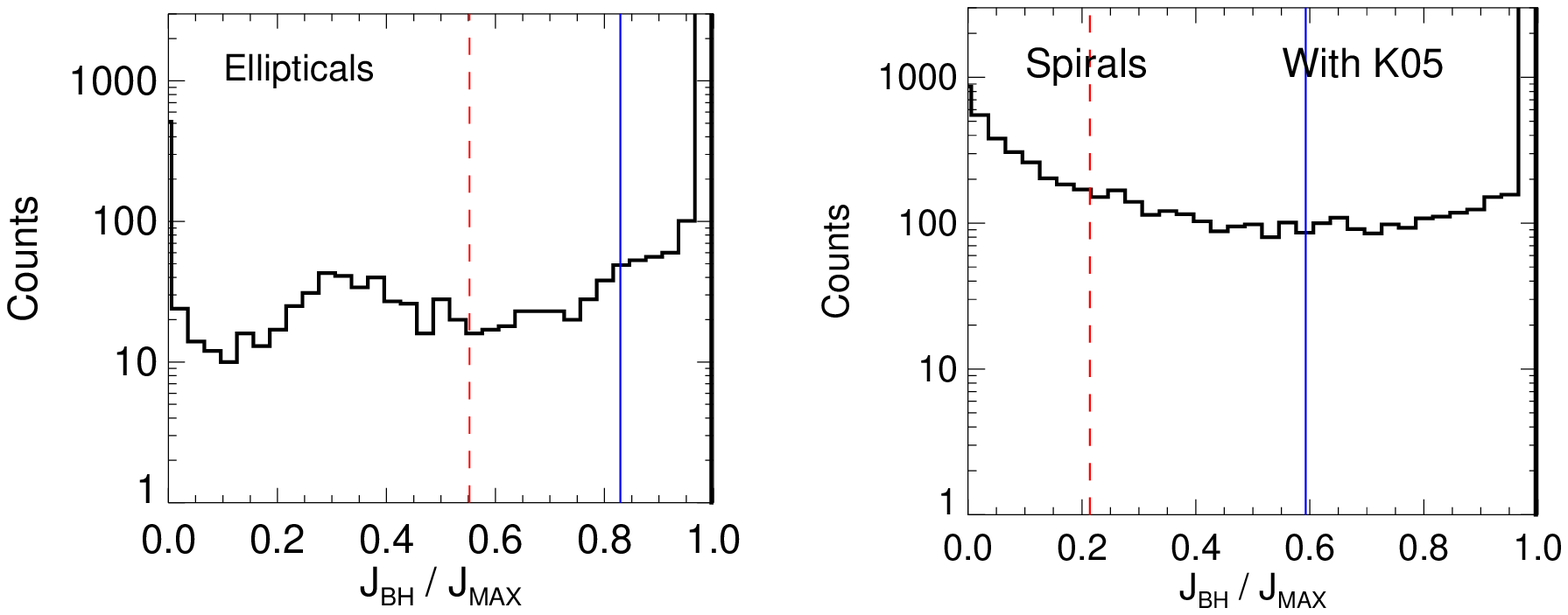}
\includegraphics[width=0.48\textwidth]{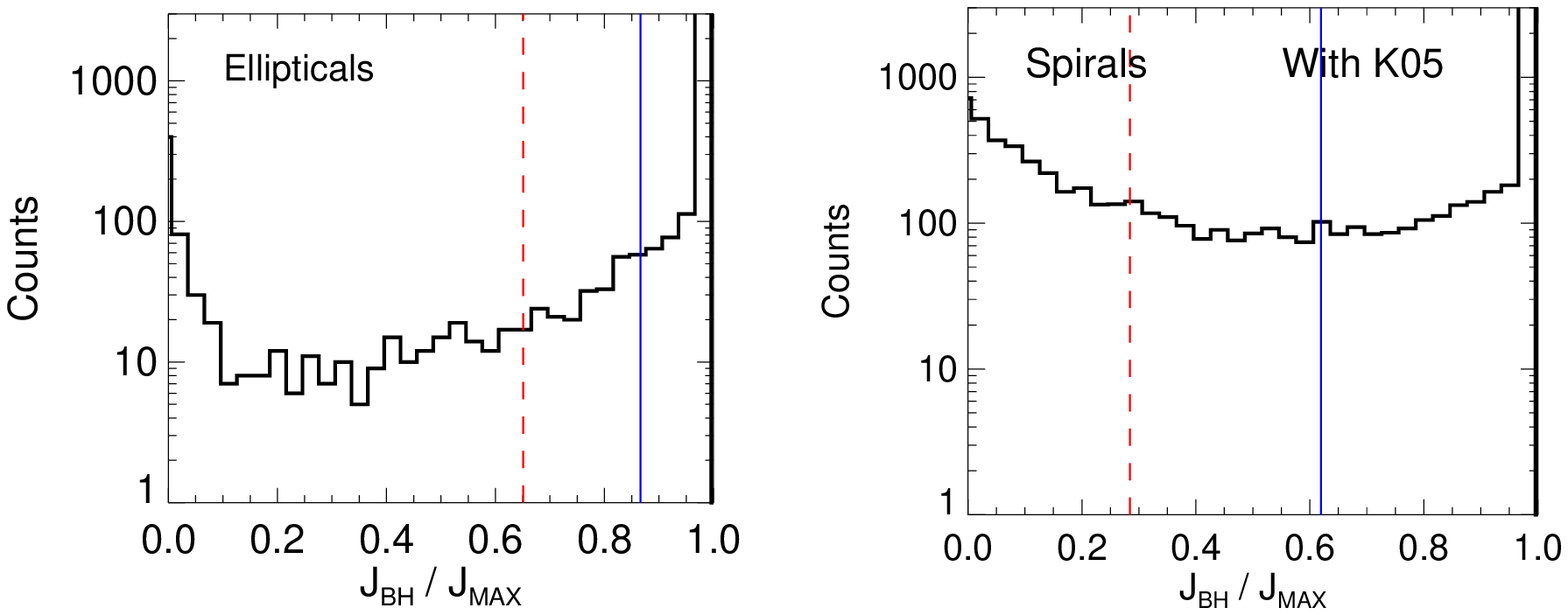}
\includegraphics[width=0.48\textwidth]{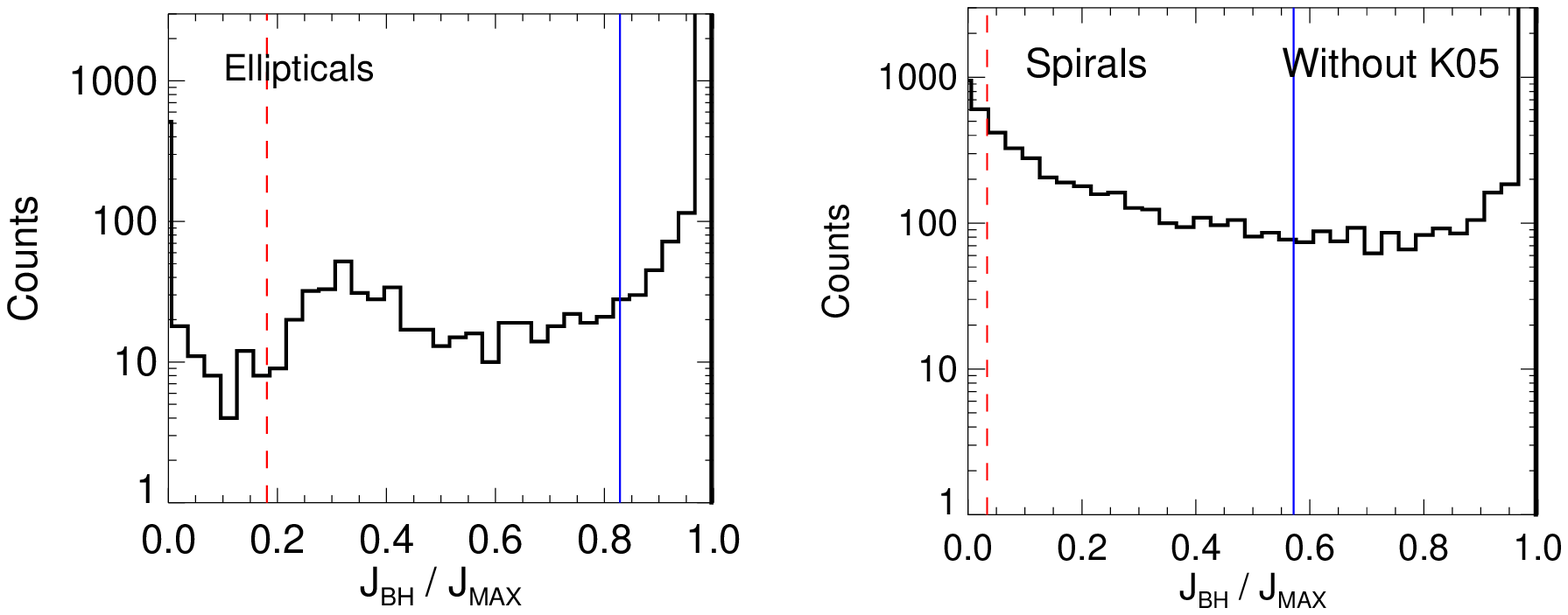}
\caption{Distributions of black hole spins at $z=0$ hosted by
elliptical (left panels) and spiral galaxies (right panels)
from model A, in which the direction of the accretion disc angular momentum 
is inferred from the 
process responsible for the gas accretion onto the central BH.
The distributions are obtained from a
constant initial BH spin
\^a$_{\rm initial}=0.01$ and the K05 alignment criterion (upper panels), 
from \^a$_{\rm initial}=M_{\rm BH, initial}/10^{7} M_{\odot}$ and K05 
(middle panels), 
and from \^a$_{\rm initial}=0.01$ and 
by adding the angular momentum of
the accretion disc to the BH spin without considering 
the K05 criterion (lower panels).
The vertical solid lines represent 
the average values of BH spin obtained from the histograms.
For reference, 
the average values of BH spin obtained
from randomly oriented accretion discs  
(model B) are shown as vertical dashed lines.}
\label{spins}
\end{center}
\end{figure}

Fig.~\ref{spins} shows the BH spin distribution for BHs hosted by elliptical
(left panels) and spiral (right panels) galaxies
for three different versions of model A, (i) 
using a constant initial BH spin \^a$_{\rm initial}=0.01$ 
and the K05 alignment criterion
(upper panels), (ii) also with K05 and considering
an initial BH spin proportional to the BH mass, 
\^a$_{\rm initial}=M_{\rm BH, initial}/10^{7} M_{\odot}$ 
(middle panels), and 
(iii) adopting the same initial BH spin as in (i) but
without considering K05  
(lower panels).
Vertical solid lines represent the average BH
spin using the full sample for each morphological type. 
For reference, average values from model B { are shown by the dashed lines}.

In all cases considered, elliptical galaxies host  
both rapidly and slowly-rotating BHs 
(the distributions show two peaks, at \^a$\, \sim 0$ and 
\^a$\, \sim 1$),  
with a marked preference for
high rotation; 
low values of spin are associated to low stellar mass galaxies
($M_{\rm Stellar} \la 10^{10} \,M_{\odot}$), 
with BH growth histories characterised by a typically low number of accretion episodes.
Spiral galaxies, on the other hand, show a larger population of low spin BHs.
Even though spirals also show a peak at \^a$\,\sim 1$, the
average values of BH spin in ellipticals, 
for both models A and B,
are always higher than the average BH spin in
spirals. { This trend does not depend on the morphology criterion, since threshold
values $r_{\rm thresh} \ga 0.7$ produce practically the same spin distributions (i.e. variations in
the average values are smaller than $20\%$). Moreover, even a threshold
as small as $r_{\rm thresh}=0.5$ produces the same trend of increasing spin values for BHs hosted 
by elliptical galaxies.}
Our results change only slightly 
when the model does not consider the K05 criterion (lower panels),
producing 
an almost negligible decrease of the average BH spin for
both elliptical and spiral galaxies. 
{ However, the effect of not applying the criterion K05
has strong impact in model B,
producing very low spin values in all populations (i.e.
accretion
episodes randomly add/subtract angular momentum to the BH).}

\begin{figure}
\begin{center}
\includegraphics[width=0.48\textwidth]{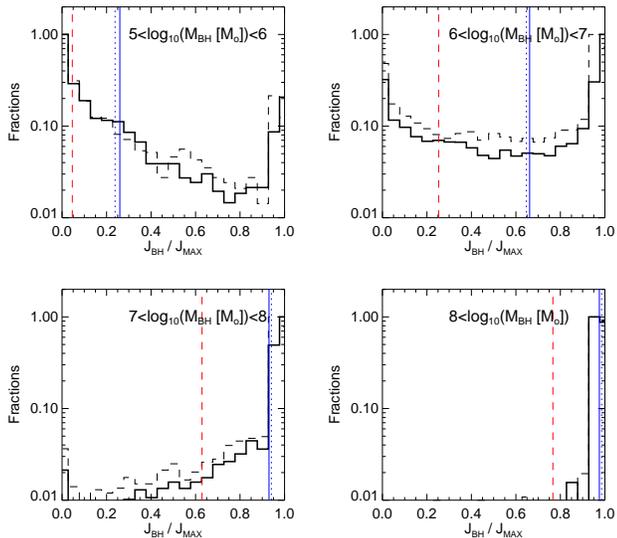}
\caption{Normalized distributions of BH spin in four different BH
mass bins obtained from model A
using a constant initial BH spin \^a$_{\rm initial}=0.01$ and the K05 alignment criterion.
Solid vertical 
lines indicate average values of the BH spin calculated
from the distributions shown
in each panel.
For reference, we also show the average values obtained from model A without K05 (dotted lines) 
and model B with K05 (dashed lines). { The dashed histogram shows the distributions
of model A with K05 and initially spinless BHs.}}
\label{HBH}
\end{center}
\end{figure}

This morphology-related dichotomy also arises from
the analysis of BH spin distributions as a function of the BH mass.
Fig.~\ref{HBH} shows the resulting normalized distributions 
from model A with an initial BH spin $\hat{a}_{\rm initial}=0.01$ and K05 alignments 
for four different BH mass ranges.
The distributions are characterised by a progressive increase of the relative number of 
BHs with high spin values as we move to higher BH masses.
This is clearly shown by the average values of BH spins
represented by the vertical solid lines.
{ Taking into account that our model reproduces the well known 
BH-bulge mass relation 
(e.g., \citealt{Magorrian98}, \citealt{Gebhardt00}, \citealt{Haring04}),
the link between massive elliptical galaxies and high BH spins is also recovered
(see Fig. \ref{spins})}.  
The similar average values denoted by solid (model A with K05) 
and dotted (model A without K05) lines
support our previous conclusion about the 
minor influence of the K05 alignment criterion on the BH spins.
{Although the average values from model B 
(random angles with K05; dashed lines) are different from those
given by model A, the general behaviours of the average values 
are similar in both models.
This is opposite to previous results by
\citealt{King08}, where the most massive objects 
with BH spins obtained by considering random angles
show the inverse trend.
The distributions for model A with K05 and with a null initial BH 
spin are shown as dashed histograms and, as can be seen, 
there are little differences with respect to the non-zero spin case (solid histogram),
and the increase in the average spin
value as a function of BH mass is also obtained.}

The three different scenarios explored here
imply that massive ellipticals host more rapidly rotating BHs than
spirals, as \citet{Sikora07} proposed in their revised
version of the ``spin paradigm''. {This fact, together with the lack of a dependence of the
$z=0$ BH spin
distributions on their initial value, 
confirm
that the accretion and assembly history 
almost entirely define the final BH spin value. This was 
suggested earlier by our analysis of the BH growth tracks
(Fig.~\ref{tracks}).}
This result is contrary to what is
found from randomly oriented accretion discs, since in this case the final
distribution retains the memory of its
initial conditions, as has already been reported in previous works (e.g.
\citealt{Volonteri05}).
We also notice that the K05 criterion has a small impact on the final
average value of the BH spin.
In the following section we discuss
into more detail the effects of considering K05 by
using an individual example galaxy extracted from the model. 

\section{Black hole spin development}\label{growth}

In this section we present in detail the
assessment of the 
contribution from each BH growth mechanism 
to the final spin value. We also evaluate the influence
of assumptions regarding the accretion disc orientation
and the effect of K05 alignments. 

\begin{figure*}
\begin{center}
\includegraphics[width=0.33\textwidth]{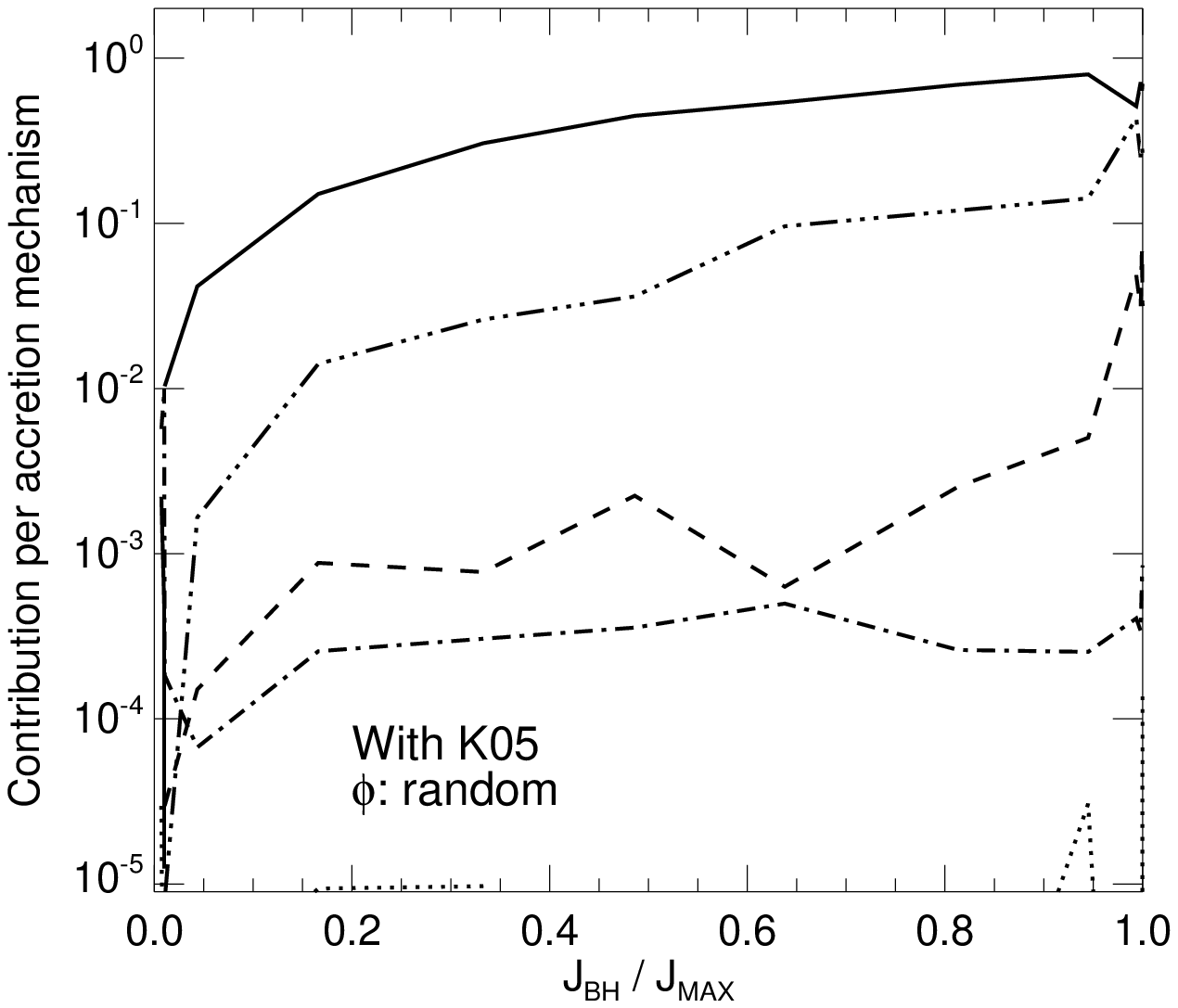}
\includegraphics[width=0.33\textwidth]{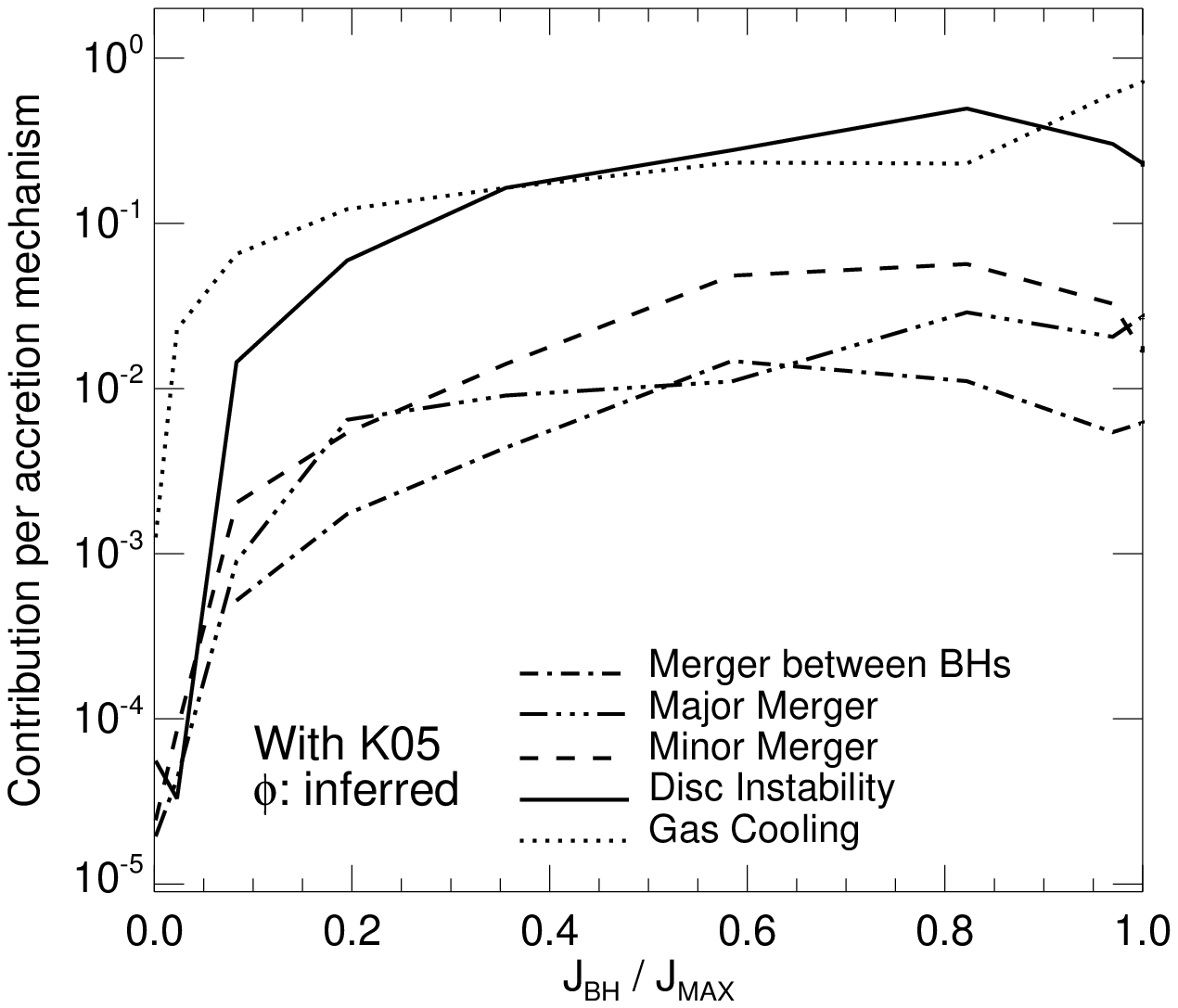}
\includegraphics[width=0.33\textwidth]{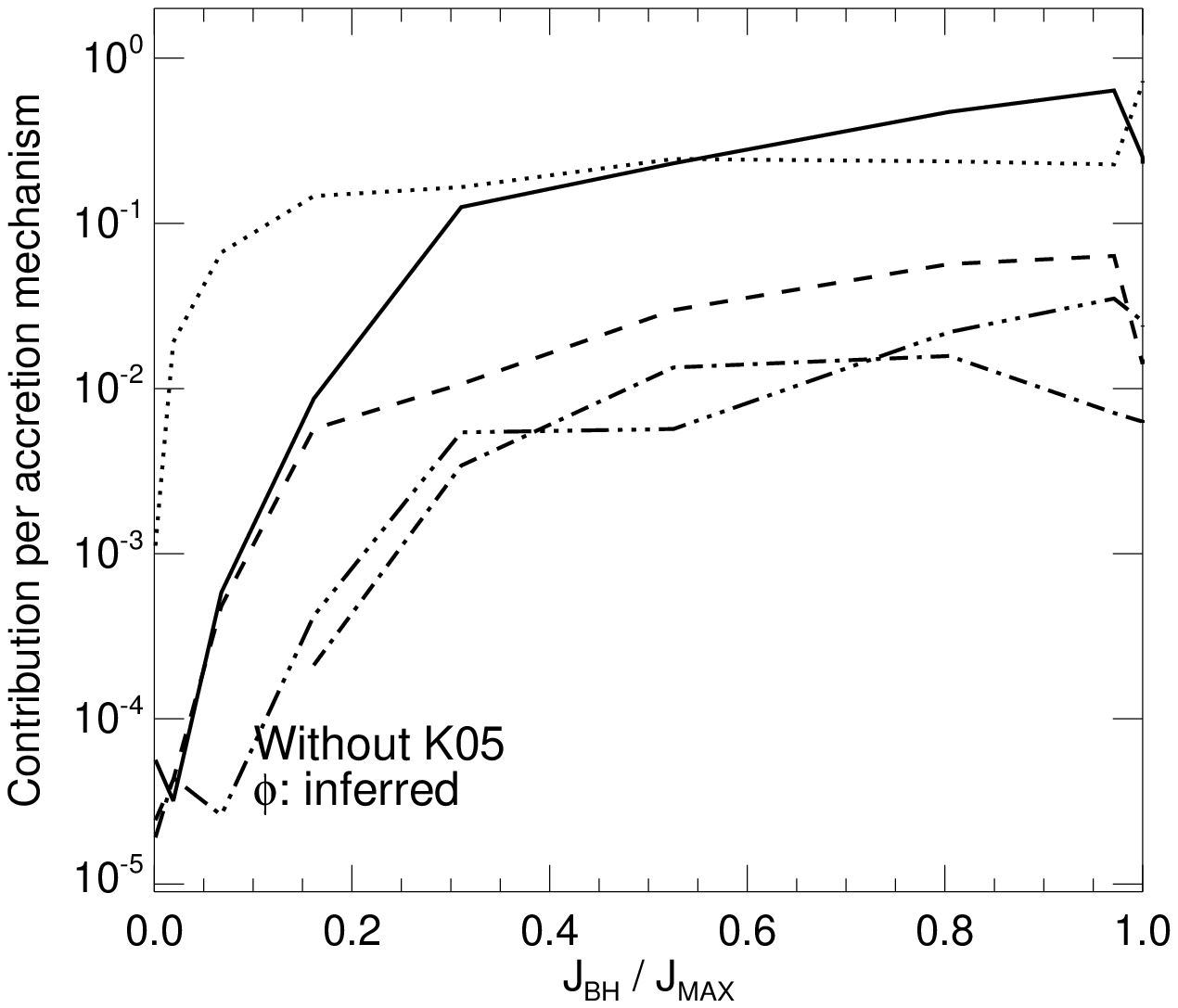}
\caption{
Contribution
to the final BH spin from the five different mechanisms
involved in the BH growth:
mergers between BHs (dot-dashed line), and gas accretion driven by
galaxy major mergers (triple dot-dashed lines),
galaxy minor mergers (dashed line), galaxy disc
instabilities (solid line), and  gas cooling (dotted lines).
\textit{Left panel:}
results from model B with a constant initial BH spin \^a$_{\rm initial}=0.01$
and the K05 alignment criterion.
\textit{Middle panel:} results from model A, with \^a$_{\rm initial}=0.01$ and K05.
\textit{Right panel:} same as middle panel but without the K05 criterion.}
\label{con1}
\end{center}
\end{figure*}

Fig.~\ref{con1} shows the contribution of each BH growth 
mechanism to the BH
spin value at $z=0$ including mergers between BHs, 
and gas accretion driven by gas 
cooling processes, 
galaxy minor/major
mergers, and disc instabilities in the host galaxy. 
The calculation of the contributions from each mechanism is explained 
in detail in
Section~\ref{model}.
The left panel shows the results obtained from
model B with a constant initial BH spin \^a$_{\rm initial}=0.01$
and the K05 alignment criterion.
We can see that the main spin-up mechanisms are 
disc instabilities and galaxy major mergers, the same processes 
that provide the main source of BH growth (Fig.~$2$ in LCP08).
This result is
not necessarily expected a priori since, in model B, the gas involved in
minor and major mergers
comes from different
galaxies with random galaxy disc orientations,
which may originate a
misalignment of the accretion disc. 
{Note that gas cooling processes produce
a minor average contribution throughout the whole BH spin range, mainly due
to the low mass accretion rates during this process in addition to the random relative orientations,
which result in roughly alternating additions and subtractions to the BH spin 
modulus.}

The other two plots of
Fig.~\ref{con1}
show results from model A, with and without considering K05 
alignments (middle and right panels, respectively). 
In both cases, we set a constant
initial BH spin \^a$_{\rm initial}=0.01$.
We can see that, as in model B (left panel), 
disc instabilities are also one of the 
main spin-up 
mechanisms. However, the main difference with respect to the
case of randomly oriented accretion discs is the important
influence gained by gas cooling processes. 
The contribution of minor mergers also becomes more important than
in model B, being comparable to that of major mergers. 

These results can be understood recalling that, during galaxy mergers, the
angular momentum of the accretion disc arises as a result
of the contribution from the gas mass-weighted angular momenta of
the merging galaxies,
while in self-interaction processes, like
gas cooling and disc instabilities, the accretion disc simply takes the
{direction of the host halo angular momentum or the galaxy disc, respectively}.
This indicates that, at high redshifts,
the first non-aligned
accretion discs are likely to come from galaxy mergers.
Taking into account the comparable influence of disc instabilities 
and gas cooling processes on the results of
model A,
our conclusions on the development of the BH spin based on this model
would not be strongly affected by the current modeling of disc instabilities,
which 
is a simplified version of the actual process taking
place in real galaxies \citep{athana08}.
This is not necessarily true for model B, where disc instabilities
are major contributors to the BH spin.

Finally, mergers between BHs make a low relative contributions to the
final value of the BH spin in all the three cases analysed.
This is in
agreement with previous works that indicate that high spins are
achieved by
accretion of baryonic matter rather than by mergers between BHs
(e.g.
\citealt{Hughes03}; \citealt{Peirani08}; \citealt{Berti08};
\citealt{Moderski96}; \citealt{Moderski98}). 
We remark that the relative importance of each growth mechanism is not
affected by changes in the initial BH spin values tested in this work.

\begin{figure*}
\begin{center}
\includegraphics[width=0.36\textwidth]{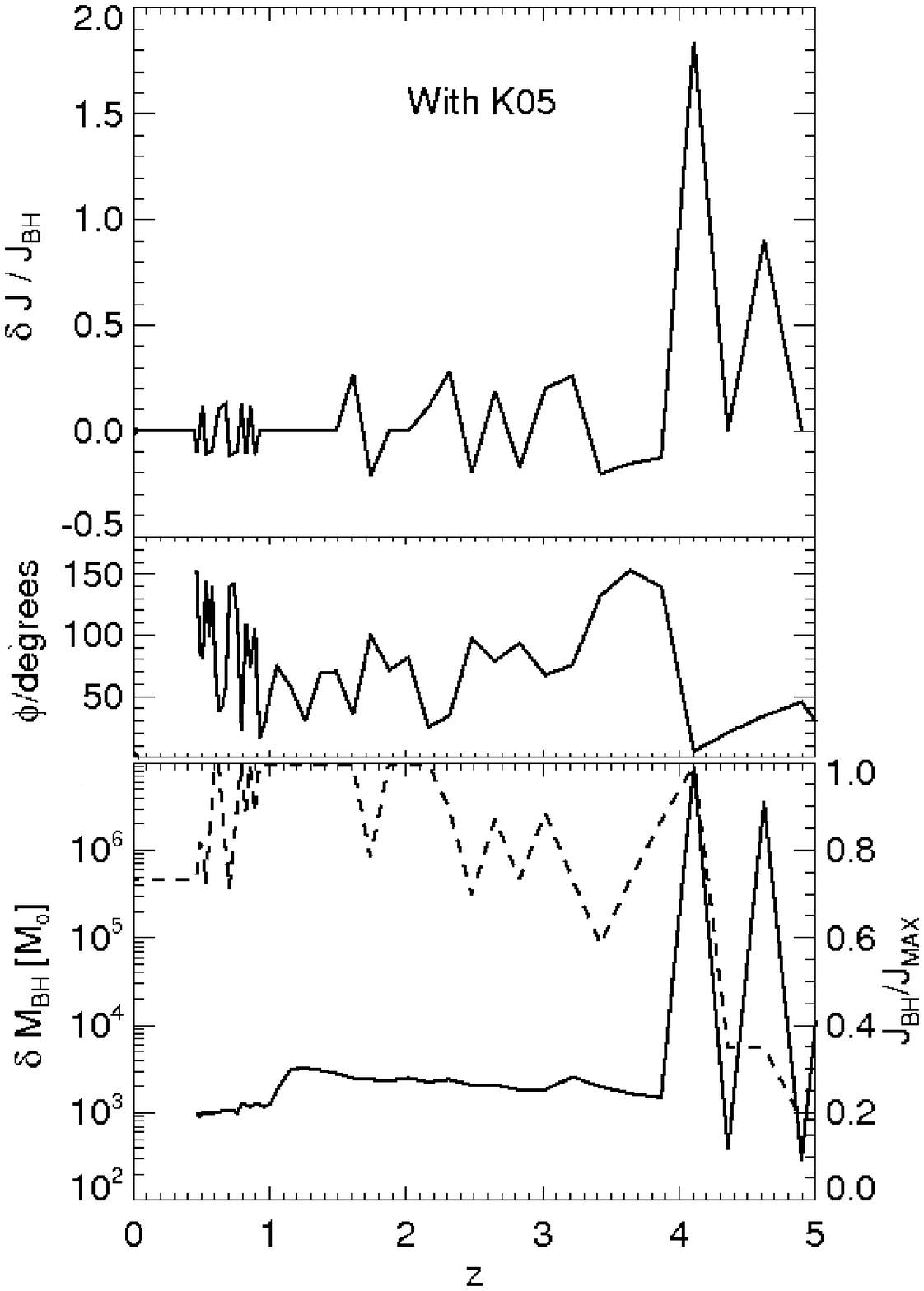}
\includegraphics[width=0.36\textwidth]{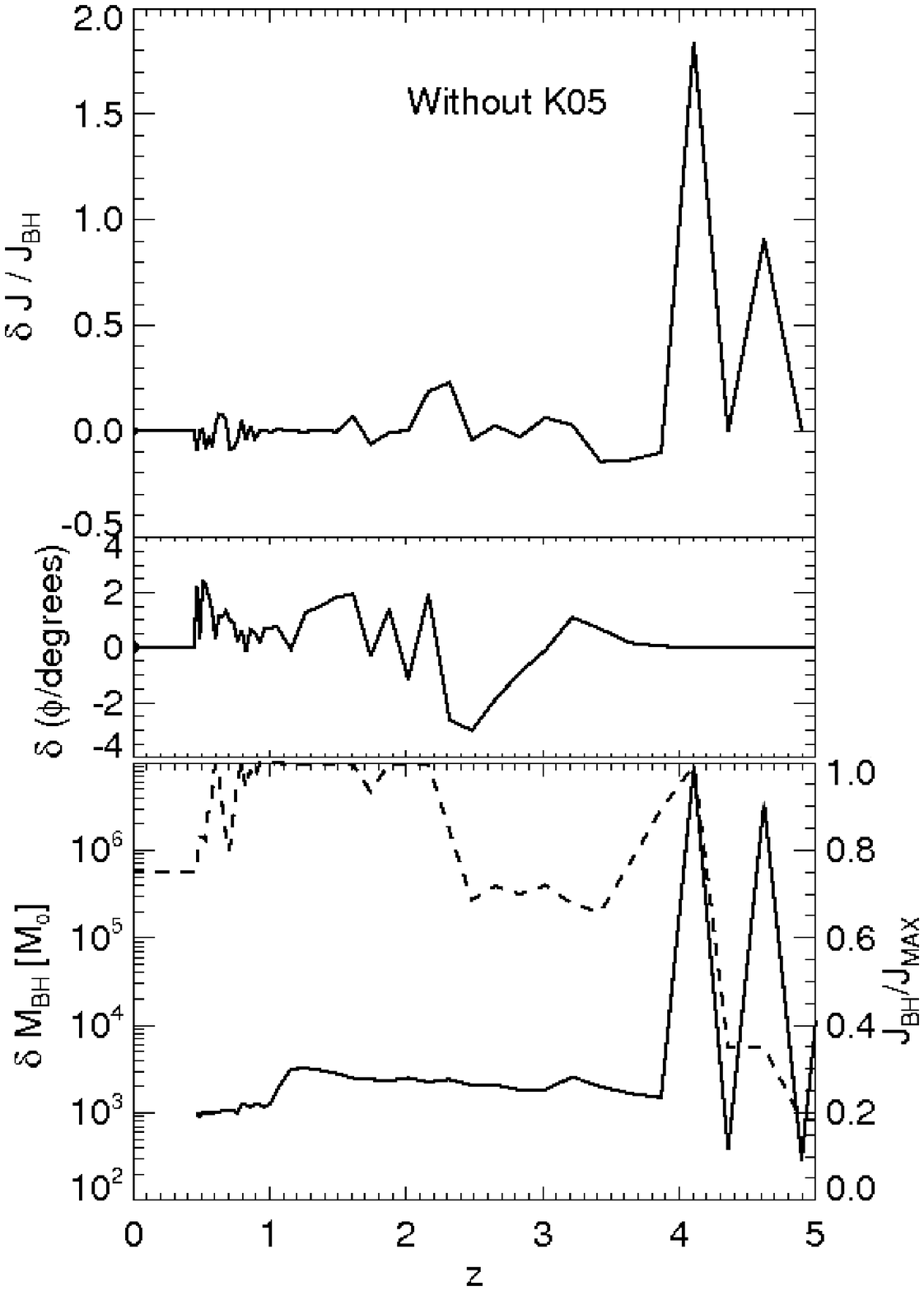}
\caption{
Redshift evolution of properties of the BH-disc system  
for an example galaxy extracted from SAG
(spiral galaxy with $z=0$ stellar mass
$M_{\rm Stellar}=8 \times 10^{9}M_{\odot}$ and a BH mass of $M_{\rm BH}=2
\times 10^{7}M_{\odot}$ ). 
The left and right panels show the results 
from model A with and without the K05 alignment criterion,
respectively.
The upper panels show the relative contribution of the angular
momentum of the accretion discs
to the angular momentum of the BH before the accretion,
$J_{\rm BH}\equiv J_{\rm BH}^{\rm initial}$, giving a variation
$\delta J=J_{\rm BH}^{\rm final}-J_{\rm BH}$
between consecutive snapshots
of the {\em N}-Body simulation.
{The middle left panel presents the angle
between the accretion disc and the BH spin, $\phi$; the middle right panel 
shows the variation in $\phi$ when the K05 criterion is removed}. 
Lower panels show the {mass accreted by the BH} 
and the BH spin development in solid and dashed lines, respectively.}
\label{phiDisk}
\end{center}
\end{figure*}

These results are only slightly modified when K05 alignments
are not used,
in which case the angular momentum of the disc
is simply vectorially added to the BH spin
without considering alignments
(right panel of Fig.~\ref{con1}).
In this case,
gas cooling processes and disc instabilities
contribute
to the BH spin
in a rather similar way to the case where the K05 criterion is considered.
This indicates that
the angular momentum of accretion discs is mainly distributed
within narrow cones around the BH spin direction, implying
a slightly increased final spin value
for a model with K05. In the more rare
case of counter-alignments, the accretion
episodes can also produce more important spin-down effects.

Regarding the effect of K05 alignments, we focus on 
an individual case in order to illustrate its effect on the BH spin development. 
Fig.~\ref{phiDisk} shows
the evolution of several properties of the
BH-disc system of one example galaxy taken from model A,
with and without considering
the K05 alignment criterion (left and right panels, respectively).
The upper panels show the relative contribution
to the angular
momentum of the BH, $\delta J/J_{\rm BH}$ {(where $\delta J=J_{\rm BH}^{\rm final}-J_{\rm
BH}^{\rm initial}$ is the angular momentum variation 
between consecutive snapshots of the underlying {\em N}-Body simulation and
$J_{\rm BH}=J_{\rm BH}^{\rm initial}$)},
coming from accretion discs formed by the different gas accretion
events (see Fig.~\ref{con1}).
In both cases, with and without K05, 
some accretion episodes contribute to the BH angular momentum in
amounts 
comparable or even higher than the current value of $J_{\rm BH}$.
{The middle left panel shows
the angle 
between the angular momentum of the accretion disc and the BH spin;
the middle right panel presents the variation in this angle arising from
not including the K05 criterion.
Lower panels depict the mass accreted by the BH in each snapshot (left-hand
y-axis) and the BH spin history (right-hand 
y-axis).
Notice that 
the
inclusion of K05
induces important changes in the details of the BH spin development.
In particular, 
the accretion episodes that take place between $2.5 \le z\le
3.5$ do not change significantly the BH spin
value when K05 is not applied, whereas when considering K05 
the BH spin shows important variations.
However, there is no final important net difference between
these two models, since the sequence of accretion episodes that increase/decrease
the BH spin almost cancel each other completely.
}
Note that applying K05 also has a small effect on the
resulting angles between BH spins and the galaxy angular momenta, producing differences
of less than $4$ degrees. 
Finally, 
it is interesting to notice, by looking at the middle panels, that in $\sim 60$ per cent of the accretion
episodes the angle between the angular momenta of the disc and the BH 
is $\phi<90$ degrees.

Our inclusion of accretion disc orientations arising from the LSS,
with or without K05 alignments,
produces important changes in the contributions
of different processes to the
BH spin development, with respect to the results
obtained using random
orientations.
This could have an important impact in our understanding of the
connection between 
the BH spin and the luminous activity of AGN.
For instance, recent studies by
\citet{Ciotti07} show that
RL QSOs might be produced by recycled gas rather than mergers. In our
model, these objects would be characterised by
high spin values (see for instance Fig.~\ref{con1}); however,
if we considered a Monte-Carlo model, these objects
would probably show a low spin
(e.g. \citealt{King07}), leading to a completely different
interpretation of the physical phenomenon at work.

\section{Behind the nature of AGN radio loudness}\label{radioloud}

The results of our model have shown that the properties
of black holes and their host galaxies are closely connected,
in agreement with numerous 
observational and 
theoretical works on this subject.
However, our model does not provide direct information
on the radio loudness of the galaxy population.
In an attempt to classify a galaxy as a radio-loud object,
we explore different possibilities using
available properties of galaxies and their BHs
provided by the model.
Among the most important quantities required to fully characterise the 
BH and its accretion disc 
are the way in which gas accretion proceeds (extensively discussed in
Section~\ref{Introsec}),
the BH mass, the BH spin,
and the magnetic field. 

\begin{figure*}
\begin{center}
\includegraphics[width=0.45\textwidth]{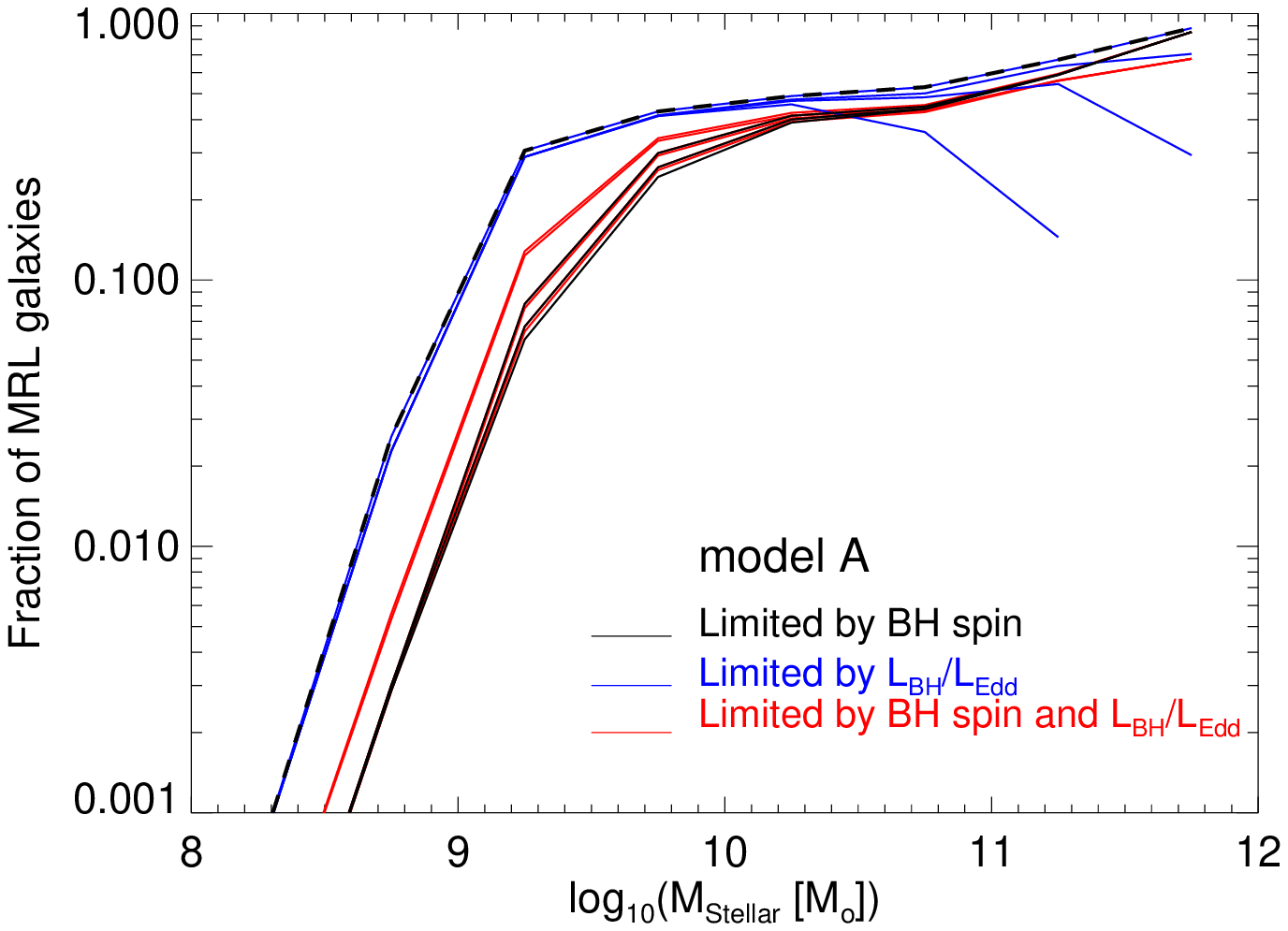}
\includegraphics[width=0.45\textwidth]{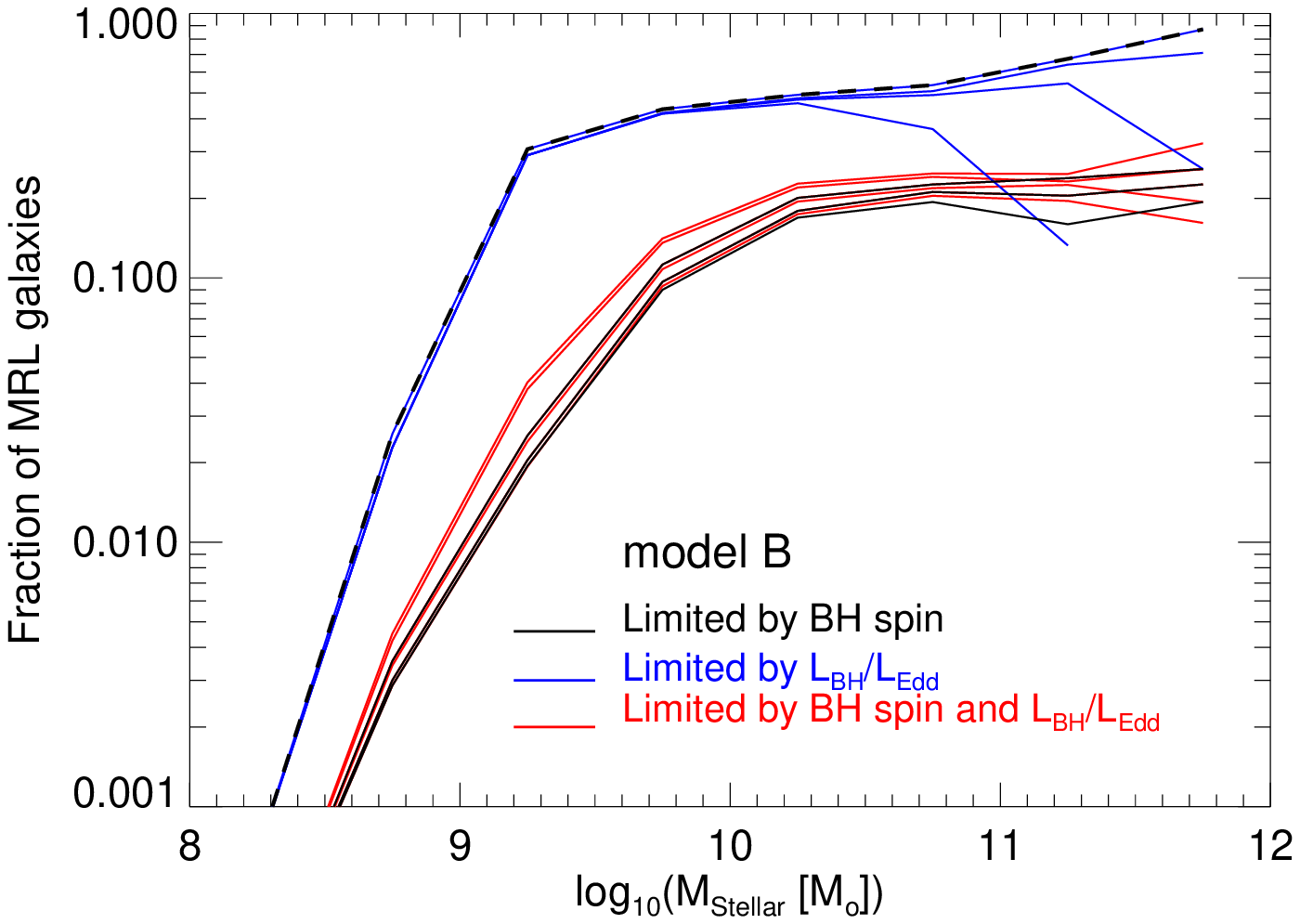}
\caption{Fraction of $z<0.1$ MRL galaxies as a function of
stellar mass for model A (left panel) and B (right panel), 
both using an initial
BH spin $\hat{a}_{\rm initial}=0.01$ and including K05 alignments. 
Blue lines show the fractions for samples with Eddington ratios
$\lambda<\lambda_{\rm MAX}$, with $\lambda_{\rm MAX}=
10^{-1}$, $10^{-2}$,
$10^{-3}$ and $10^{-4}$ (top to bottom lines); 
black lines, the
fractions for samples with BH spin
$\hat{a}>\hat{a}_{\rm MIN}$, with
$\hat{a}_{\rm MIN}=0.3$, $0.6$ and $0.9$ (top to bottom lines);
red lines show the
fractions  
resulting from combinations of limits on both accretion rate and BH spin.
For reference, 
we show the 
fraction of active galaxies 
as a function of stellar mass (black dashed lines). 
}
\label{RadioStellar}
\end{center}
\end{figure*}

Observational results indicate that the average BH mass in
radio-loud systems is about twice as large as  
in radio-quiet galaxies
(e.g. \citealt{Corbin97}, \citealt{Laor00}, \citealt{Lacy01},
\citealt{Boroson02}, \citealt{Jarvis02}, \citealt{Metcalf06}).
Taking into account
the well known
relationship between BH and host bulge mass, it is
natural to extend
the dependence of radio loudness 
to the stellar mass of the host galaxy.
Nonetheless, 
both populations of RL and
RQ QSOs have been detected in massive elliptical galaxies
(e.g. \citealt{Floyd04}); also, BHs of equal mass
show differing radio luminosities by 
up to $\sim 4$ orders of magnitude 
(\citealt{Ho02}, \citealt{Dunlop03}, \citealt{Sikora07}).
Therefore, it can be inferred that 
the influence of this BH property 
on the jet production, and the consequent strength of the radio emission,
is secondary. Other results in this direction are found by 
\citet{Merloni03}.

The second parameter that may play a role in the RL/RQ dichotomy is the
amplitude of the local magnetic field from accretion flows which, in
some cases, can 
produce large-scale fields that would help 
collimate 
strong relativistic jets
\citep{Krolik07}.  It has been argued, however, that the presence of 
such fields 
may not be of primary importance \citep{spruit}.
In microquasars, that 
share the same physical principles than AGN
on much smaller scales, 
it is possible that cold winds are responsible for the collimation of jets
(\citealt{Bosch06}).  This alternative 
further diminishes the potential relevance of magnetic fields as
a {primary} collimation mechanism.

The third physical property 
that can be linked to the   
radio loudness phenomenon is
the BH spin as has been suggested by several authors
(e.g. \citealt{Further95},
\citealt{Lacy01}; \citealt{Lacy02}; \citealt{Hawley06}; \citealt{Ballantyne07}; 
\citealt{Nemmen07}; \citealt{Cen07};
\citealt{Sikora07}; \citealt{Volonteri07}). Furthermore, a recent study by 
\citet{Lindner07} shows that the jet direction is influenced by the BH spin.
Therefore, in addition to the accretion rate,
we consider the BH spin as another possible candidate
to break the degeneracy between the RQ and RL populations.

With these caveats in mind, 
we assume that 
radio loudness is defined by the 
BH spin, the
gas accretion rate (``accretion paradigm''), or combinations of both parameters
(as in the ``revised spin paradigm'', \citealt{Sikora07}). 
We evaluate the possibility that
radio loudness is defined by these criteria using 
observational results on the fraction of radio-loud
galaxies ($f_{\rm RL}$) which is thought to increase with 
both, BH and stellar mass
(\citealt{Corbin97}; \citealt{Laor00}; \citealt{Lacy01};
\citealt{Boroson02}; \citealt{Jarvis02}; \citealt{Best05};
\citealt{Metcalf06}). { These relations are still a matter of debate,
since it has been argued that they can simply be the result of observational
biases (see for instance \citealt{White07}).  In addition, since there are no
measurements of $f_{\rm RL}$ using complete
samples of galaxies to date, we are limited to make qualitative
comparisons to the observed data.  The most complete study performed to date was
presented by
\citet{Best05} based on the Sloan Digital Sky Survey (SDSS) but, even in this
case, they only consider FR Class 1
radio sources, characterised by low radio luminosities. 
We will attempt
to distinguish between the predictions of models A and B
by using only qualitative comparisons.
}

To this aim, we focus on $z\approx 0$ active galaxies (i.e.
with Eddington ratios $\lambda\equiv L_{\rm BH}/L_{\rm Edd}>0$), and
separate them into samples of model radio-loud galaxies (MLR)
selected using different upper limits in accretion rates,
$\lambda <\lambda_{\rm MAX}$, 
with $\lambda_{\rm MAX}=
10^{-1}$, $10^{-2}$,
$10^{-3}$ and $10^{-4}$, and three normalized minimum BH spin values, 
$\hat{a}_{\rm MIN}=0.3$, $0.6$
and $0.9$.  We test whether these MRL samples
are able to mimic the observed radio-loud population.
Fig.~\ref{RadioStellar} shows
the resulting fractions of MRL galaxies, $f_{\rm MRL}$, 
as a function of stellar mass
for models A and B (cf. Section~\ref{advantage})
using an initial 
BH spin $\hat{a}_{\rm initial}=0.01$ and including K05 alignments. 
Fractions obtained using thresholds in Eddington ratio only,
$\lambda < \lambda_{\rm MAX}$,  are represented
by blue lines.
Black lines show the results for galaxies
selected using $\hat{a}>\hat{a}_{\rm MIN}$, and
red lines, the results for samples obtained from the combination of limits
on both spins and accretion rates. 
For reference, the results for the 
full population of active galaxies are represented
by a black dashed line.

As can be seen, for model A,
the limits on Eddington ratios produce 
very similar fractions to what is obtained for
the full AGN sample, mostly
due to the lack of BHs with high $\lambda$ values at low redshift.
The main features for samples limited by $\lambda$ are
an increasing $f_{\rm MRL}$ with stellar mass for 
$M_{\rm Stellar} \la 10^{9} M_{\odot}$, with a plateau up to
$M_{\rm Stellar} \approx 10^{10} M_{\odot}$.
Stringent limits,
$\lambda_{\rm MAX} \la 10^{-3}$,
produce a decreasing
$f_{\rm MRL}$ at the high stellar mass end.  
However these limits are very restrictive since 
observed RL galaxies can show
Eddington ratios as large as $\lambda \approx 10^{-2}$
(e.g. \citealt{Ho02}; \citealt{Sikora07}).
This aspect should be taken into account in order
to test these particular predictions.  
These trends are at odds
with the monotonically increasing observational relation mentioned above.  There
is little effect on these trends when considering the case of a null initial BH spin.

On the other hand, when considering
thresholds on the BH spin alone, the fractions of MRL galaxies
always increase with stellar mass.
The results for the combined conditions, $\hat{a}>\hat{a}_{\rm MIN}$ and
$\lambda<\lambda_{\rm MAX}$, are only shown for 
$\lambda_{\rm MAX}=10^{-1}$ and $10^{-2}$.
These cases also show increasing relations between $f_{\rm MRL}$
and stellar mass because of the strong influence
of the limits on the BH spin.
Finally, for all the selection criteria analysed,
model B (right panel of Fig.~\ref{RadioStellar}), 
characterised by random orientations, 
fails to produce monotonically increasing fractions of MLR galaxies
as a function of stellar mass.
{When we consider null initial BH spins, the only 
appreciable change in the fractions of 
MLR galaxies is a slight increase in the slope of the relation
with respect to the results shown in Fig.~\ref{RadioStellar},
}

\begin{figure}
\begin{center}
\includegraphics[width=0.45\textwidth]{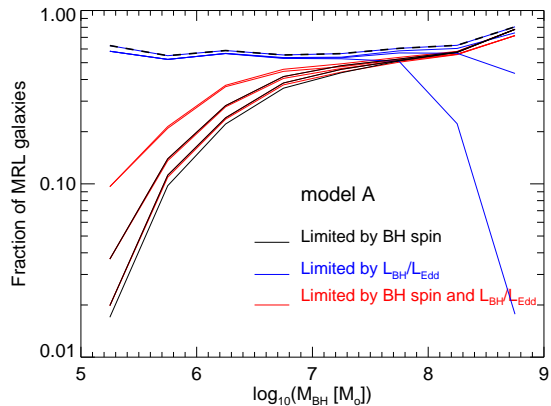}
\caption{Fraction of MRL galaxies as a function of
BH mass for model A using an initial
BH spin $\hat{a}_{\rm initial}=0.01$ and including K05 alignments.
Line types and galaxy samples are as in
Fig.~\ref{RadioStellar}.}
\label{RadioBH}
\end{center}
\end{figure}

Fig.~\ref{RadioBH} shows the fractions of MRL galaxies 
as a function of BH mass for model A 
(lines colours are as in Fig.~\ref{RadioStellar}).
As can be seen, the fractions of galaxies selected according to the
three sets of criteria on BH spin and accretion rates
show similar behaviours 
to those obtained as a function of stellar mass. 
The same similarity between the dependences with BH and stellar mass
arise for model B, with milder or negligible trends of the MRL 
fractions with BH mass.

All the previous results correspond to the local Universe.
From our results on the development of BH spin (see for instance Fig.~\ref{phiDisk}), it is
natural to expect smaller fractions $f_{\rm MRL}$ at higher redshifts,
particularly when using the BH spin as the main parameter to
define radio-loudness. 
However, the fractions of active galaxies increase
importantly at higher redshifts producing a combined effect.
Galaxies hosting BHs with masses
$M_{\rm BH} \ga 10^8 M_{\odot}$ lead to higher fractions $f_{\rm MRL}$ with
respect to those with
BHs of the same mass in the local Universe, while galaxies with
BHs masses in the range $M_{\rm BH} \la
10
^8
M_{\odot}$ show the expected decreased fractions.
As the redshift
increases the relation becomes steeper as a function of both BH and stellar mass.
This prediction can be confirmed by further observations of high
redshift radio-sources.

We have combined two models for the orientation of accretion discs with
three sets of criteria related to the accretion and spin paradigms, with the aim
of isolating the main parameters that characterise the radio loudness of
galaxies.
Our results indicate that the dependence of the fractions of MRL galaxies
with BH and stellar mass are a qualitative match to
the increasing trend showed by observational results
when thresholds on the
BH spin 
are applied. It is important to notice that this is 
more naturally achieved in a model
with LSS oriented accretion discs.
The most important aspect to emphasize from our results is that this qualitative
agreement preferentially favours the ``spin paradigm''
as the likely scenario for radio loudness in the Universe. We remind the reader
that the observed trend
is still in
debate. Therefore, 
further observational data on AGN are required
in order to reach a proper understanding of
the biases involved in the measurements of fractions
of RL galaxies. Only once this problem is solved it will be possible
to attempt to tip the scales in favour a single model, for example A or B. 

\section{Summary and perspectives}\label{conclusion}

We have implemented a simple
calculation of the BH spin 
in the semi-analytic model of galaxy formation
described by LCP08 to study relationships between 
the BH spin, the BH mass and the host galaxy properties. 
The main advantage of using
a semi-analytic model relies in the ability to follow, for the first time,
the development of the BH spin
from the contribution of different BH growth mechanisms,
all immersed within a $\Lambda$CDM cosmology.
Such processes are 
gas cooling, disc instabilities,  galaxy mergers and 
mergers between BHs. 
Additionally, the model allows us to relate phenomena taking 
place at very small scales (BH
and accretion disc) to those at larger scales (galaxy discs);
we infer the orientation of the accretion disc using that of
the
{source of accreted material (referred to as model A). 
Thus,
the angular momentum of the host halo is used 
when the accretion is driven by gas cooling, while that of the galaxy disc 
is considered
during disc
instability events; in the case of mergers, gas mass-weighted averages 
of the angular momentum of 
intervening galaxies are adopted.}
We also consider
random accretion disc orientations (model B),
as has been adopted in 
several previous works (e.g. \citealt{Volonteri07}, \citealt{King07}).

The main results and conclusions of this work are summarised as follows.
\begin{itemize}
\item 
Massive bulges, and therefore giant elliptical galaxies,
host almost exclusively rapidly rotating BHs.
Although this effect is more pronounced in model A than in model B,  
in both models
elliptical galaxies host, on average, BHs with higher spin values than spiral
galaxies, in accordance with the revised ``spin paradigm'' proposed by
\citet{Sikora07}.
Therefore, the morphology-related bimodality of radio loudness 
could be explained by
the BH spin.
\item 
High mass BHs 
($M_{\rm BH} > 10^7 M_{\odot}$)
in model A, show little dependence on the initial BH spin assigned to them,
indicating that their spin is almost entirely determined by the accretion
history.  
Additionally, including the K05 alignments of accretion discs only produces
very mild changes in the final BH spin distribution. 
Results from model B are found to be more dependent on the
initial values of the BH spin and strongly dependent on whether the K05 alignment
is applied or not.
\item For model A, gas cooling 
processes and disc instabilities are the main contributors
to the final BH spin.
It is important to bear in mind
that gas cooling processes produce very low relative accretion onto the BH.
On the other hand, in model B, the main mechanisms for BH spin-up are also
those responsible for the highest accretion rates, namely, galaxy mergers and
disc instabilities. This represents the main 
difference between the two schemes adopted for
the orientation of accretion discs.
\item 
We use measurements of the dependence of
the fraction of RL galaxies as a function of stellar and BH masses in order to
test different parameter combinations to select
a population
of model galaxies representative of the observed
RL galaxy population. 
These parameters are assumed to be
the accretion rate and/or the BH spin.
We test three different criteria, using (i) an upper limit on the
Eddington ratio (``accretion paradigm''), 
(ii)  a lower limit on the BH spin, and
(iii) simultaneous limits on both quantities (similar to the revised version of
the ``spin paradigm''). For model A, we find that criterion (i) predicts a
roughly constant fraction of possible radio-loud 
galaxies as a function of BH and
stellar mass (for $M_{\star}> 10^{9} M_{\odot}$), while criterion (ii)
produces a monotonically increasing relation.
Finally, criterion (iii) also produces an increasing
relation, although mostly due to the strong influence 
of the limit on the BH spin. 
For model B, we find that none of the proposed criteria are able to produce an increasing 
fraction of radio-loud galaxies with either the stellar or BH mass.
\item Taking into account recent observational results (e.g. \citealt{Best05}; 
\citealt{Chiaberge05}; \citealt{Metcalf06}) reporting an increasing fraction of
radio-loud galaxies with BH and stellar masses, we find that the
most suitable paradigm to explain radio loudness is the ``spin paradigm''
\citep{Sikora07}.
Still, it should be taken into account that these measurements
may be subject to observational biases (e.g. \citealt{Cirasuolo03},
\citealt{White07}).
\end{itemize}

The implementation of a BH spin model presented here can be used 
as a new tool to understand the
role of different processes on both very large and very small scales, within
an ab-initio model of galaxy formation. 
Within our modeling we are able to
reproduce results from Monte-Carlo attempts at 
describing BHs and the radio-loud galaxy
population.  Our analysis also singles out possible
systematic biases in the model predictions resulting from 
the use of random orientations
for the accretion discs which continuously form around BHs. 
Our finding of a possible influence
of the LSS alignments on the detailed spin-up history of BHs can now also
be considered in simple models, and be studied into more detail.
However, new and improved
observational constraints are still required
to help distinguish between possible causes for the AGN radio
activity in the Universe.

\section{Acknowledgements}
We would like to thank Andrea Merloni, Elena Gallo, Andreas 
Reisenegger, Gustavo E. Romero and Rodrigo Parra 
for useful comments
and discussions.
We acknowledge the annonymous referee for
helpful remarks
that allowed us to
improve this work.
We also thank the Scientific Organizing Committee of 
``First La Plata International School: Compact
Objects and their Emission'' for giving us the possibility to present
preliminary results of this work.
CL thanks Direcci\'on de Posgrado, PUC and project Gemini-CONICYT No. 32070011 for travel grants.
NP was supported by Fondecyt grant No. 1071006.  The
authors benefited from a visit of SC to Santiago de Chile supported by
Fondecyt grant No. 7080131. This work was supported in part by the 
FONDAP Centro de Astrof\'isica,
by
PIP 5000/2005 from Consejo Nacional de Investigaciones
Cient\'ificas y T\'ecnicas, Argentina, and PICT 26049 of Agencia
de Promoci\'on Cient\'ifica y T\'ecnica, Argentina.

\label{lastpage}


\begin{thebibliography}{99}
\bibitem[\protect\citeauthoryear{Athanassoula}{2008}]{athana08} Athanassoula E.,
2008, accepted for publication in MNRAS Letters, arXiv:0808.0016
\bibitem[\protect\citeauthoryear{Ballantyne}{2007}]{Ballantyne07} Ballantyne D.
R., 2007, MPLA, 22, 2397
\bibitem[\protect\citeauthoryear{Bardeen}{1970}]{Bardeen70} Bardeen J.M., 1970, Natur, 226, 64
\bibitem[\protect\citeauthoryear{Begelman, Volonteri \& Rees}{2006}]{Begelman06}
Begelman M.C., Volonteri M., Rees M.J., 2006, MNRAS, 370, 289
\bibitem[\protect\citeauthoryear{Bernadetta \& Volonteri}{2008}]{Bernadetta08}
Bernadetta D., Marta V., 2008, astro-ph:0810.1057
\bibitem[\protect\citeauthoryear{Berti \& Volonteri}{2008}]{Berti08} Berti E.,
Volonteri M., 2008, ApJ, 684, 822
\bibitem[\protect\citeauthoryear{Best et al.}{2005}]{Best05} Best P.N.,
Kauffmann G., Heckman T.x M., Brinchmann J., Charlot S.,
Ivezi\'c Z., White S.D.M., 2005, MNRAS, 362, 25
\bibitem[\protect\citeauthoryear{Best et al.}{2007}]{Best07} Best P.N., 2007,
NewAR, 51, 168 
\bibitem[\protect\citeauthoryear{Blandford}{1990}]{Blandford90} Blandford, 1990, in Active Galactic Nuclei, ed. T.J.-L. Courvoisier
\& M. Mayor
\bibitem[\protect\citeauthoryear{Bogdanovi\'c, Reynolds \&
Miller}{2007}]{Bogdanovic07} Bogdanovi\'c T., Reynolds C.S., Miller M.C.,
2007, ApJ, 661, 147
\bibitem[\protect\citeauthoryear{Boroson}{2002}]{Boroson02} Boroson T., 2002, ApJ, 565, 78
\bibitem[\protect\citeauthoryear{Bosch-Ramon, Romero \& Paredes}{2006}]{Bosch06}
Bosch-Ramon V., Romero G.E., Paredes J.M.,
2006, A\&A, 447, 263.
\bibitem[\protect\citeauthoryear{Bower et al.}{2006}]{Bower06} Bower R., Benson
A., Malbon R., Helly J., Frenk C., Baugh C., Cole S., Lacey C., 2006, MNRAS,
370, 645
\bibitem[\protect\citeauthoryear{Brenneman \& Reynolds}{2006}]{Brenneman06} Brenneman L. W., Reynolds C. S. 2006, ApJ, 652, 1028
\bibitem[\protect\citeauthoryear{Cattaneo et al.}{2006}]{Cattaneo06} Cattaneo
A., Dekel A., Devriendt J., Guiderdoni B., Blaizot J., 2006, MNRAS, 370, 1651
\bibitem[\protect\citeauthoryear{Cen}{2007}]{Cen07} Cen R., 2007, astro-ph:2660
\bibitem[\protect\citeauthoryear{Ciotti \& Ostriker}{2007}]{Ciotti07} Ciotti L., Ostriker J., 2007, ApJ, 665, 1038
\bibitem[\protect\citeauthoryear{Cirasuolo et al.}{2003}]{Cirasuolo03} 
Cirasuolo M., Celotti A., Magliocchetti M., Danese L., 2003, MNRAS, 346, 447
\bibitem[\protect\citeauthoryear{Cole et al.}{2000}]{Cole00} Cole S., Lacey
C.G., Baugh C.M., Frenk C.S., 2000, MNRAS, 319, 168
\bibitem[\protect\citeauthoryear{Conselice}{2006}]{Conselice06} 
Conselice C., 2006, MNRAS, 373, 1389
\bibitem[\protect\citeauthoryear{Corbin}{1997}]{Corbin97} Corbin M., 1997,
ApJS, 113, 245
\bibitem[\protect\citeauthoryear{Croom et al.}{2004}]{Croom04} Croom S.M., Smith R.J., Boyle B.J., Shanks T., Miller L., Outram P.J., Loaring N.S., 2004, MNRAS, 349, 1397
\bibitem[\protect\citeauthoryear{Croton et al.}{2006}]{Croton06} Croton D., et
al., 2006, MNRAS, 365, 11
\bibitem[\protect\citeauthoryear{Chiaberge, Capetti \& Macchetto}{2005}]{Chiaberge05} Chiaberge M., Capetti A., Macchetto F.D., 2005, ApJ, 625, 716
\bibitem[\protect\citeauthoryear{Chambers, Miley \& van Breugel}{1987}]{Chambers87} 
Chambers K.C., Miley G.K., van Breugel W., 1987, Nature, 329, 604
\bibitem[\protect\citeauthoryear{Donahue et al.}{2005}]{Donahue05} Donahue M.,
Voit G., O'Dea C., Baum S., Sparks W., 2005, ApJ, 630, 13
\bibitem[\protect\citeauthoryear{Dubus, Hameury \& Lasota}{2001}]{Dubus01} Dubus G., Hameury J.-M., Lasota J.-P. 2001, A\&A, 373, 251
\bibitem[\protect\citeauthoryear{Dunlop et al.}{2003}]{Dunlop03} 
Dunlop J.S., McLure R.J., Kukula M.J., Baum S.A., O'Dea C.P., Hughes
D.H., 2003, MNRAS, 340, 1095
\bibitem[\protect\citeauthoryear{Elmegreen, Bournaud \& Elmegreen}{2008}]{Elmegreen08} Elmegreen B.G., Bournaud
F., Elmegreen D.M., 2008, ApJ, 684, 829
\bibitem[\protect\citeauthoryear{Falke, K\"ording \& Markoff}{2004}]{Falcke04} 
Falcke H., K\"ording E., Markoff S., 2004, A\&A, 414, 895
\bibitem[\protect\citeauthoryear{Fanaroff \& Riley}{1974}]{Fanaroff74} Fanaroff B.L., Riley J.M., 1974, MNRAS, 167, 31
\bibitem[\protect\citeauthoryear{Fender, Belloni \& Gallo}{2004}]{Fender04}
Fender R.P., Belloni T.M., Gallo E., 2004, MNRAS, 355, 1105
\bibitem[\protect\citeauthoryear{Floyd et al.}{2004}]{Floyd04} Floyd D.J.E.,
Kukula M.J., Dunlop J.S., McLure R.J.,
Miller L., Percival W.J., Baum S.A., O'Dea C.P.,
2004, MNRAS, 355, 196
\bibitem[\protect\citeauthoryear{Gallo, Fender \& Pooley}{2003}]{Gallo03} Gallo
E., Fender R.P., Pooley G.G., 2003, MNRAS, 344, 60
\bibitem[\protect\citeauthoryear{Gebhardt et al.}{2000}]{Gebhardt00} Gebhardt K., Bender R., Bower G., Dressler A., Faber S.M., Filippenko A., Green
R., Grillmair C., Ho L., Kormendy J., 2000, ApJ, 539, 13
\bibitem[\protect\citeauthoryear{H\"aring \& Rix}{2004}]{Haring04} H\"aring N.,
Rix H., 2004, ApJ, 604, 89
\bibitem[\protect\citeauthoryear{Hawley \& Krolik}{2006}]{Hawley06} Hawley J.F., Krolik J.H., 2006, ApJ, 641, 103
\bibitem[\protect\citeauthoryear{Ho}{2002}]{Ho02} 
Ho L.C., 2002, Apj, 564, 120
\bibitem[\protect\citeauthoryear{Hughes \& Blanford}{2003}]{Hughes03} Hughes S.A., Blandford R.D., 2003, ApJ, 585, 101
\bibitem[\protect\citeauthoryear{Inskip et al.}{2005}]{Inskip05} Inskip K.
J., Best P.N., Longair M.S., R\"ottgering H.J.A., 2005, MNRAS, 359, 1393
\bibitem[\protect\citeauthoryear{Jarvis \& McLure}{2002}]{Jarvis02} Jarvis M.J., McLure R.J., 2002, MNRAS, 336, 38
\bibitem[\protect\citeauthoryear{Kauffmann}{2003}]{kauffman03} Kauffmann G.,
Heckman T.M., Tremonti C., Brinchmann J., Charlot S., White S.D.M., Ridgway
S.E., Brinkmann J., Fukugita M., Hall P.B., 2003, MNRAS, 346, 1055
\bibitem[\protect\citeauthoryear{King et al.}{2005}]{King05} King A.R., Lubow
S.H., Ogilvie G.I., Pringle J.E., 2005, MNRAS, 363, 49
\bibitem[\protect\citeauthoryear{King \& Pringle}{2007}]{King07} King A.R., 
Pringle J.E., 2007, MNRAS, 377, 25
\bibitem[\protect\citeauthoryear{King, Pringle \& Hofmann}{2008}]{King08} King
A.R., Pringle J.E., Hofmann J.A., 2008, MNRAS, 385, 1621
\bibitem[\protect\citeauthoryear{Kellermann et al.}{1989}]{Kellermann89}
Kellermann K.I., Sramek R., Schmidt M., Shaffer D.B., Green R.,
1989, AJ, 98, 1195
\bibitem[\protect\citeauthoryear{K\"ording, Jester \& Fender}{2006}]{Kording06}
K\"ording E.G., Jester S., Fender R., 2006, MNRAS, 372, 1366
\bibitem[\protect\citeauthoryear{Krolik \& Hawley}{2007}]{Krolik07} Krolik J.H., Hawley J., 2007, AIPC, 924, 801
\bibitem[\protect\citeauthoryear{Krolik}{1999}]{Krolik99} Krolik J.H., 1999, Active Galactic Nuclei, Princeton Series in Astrophysics, Princeton, USA.
\bibitem[\protect\citeauthoryear{Kumar \& Pringle}{1985}]{Kumar85} Kumar S., Pringle, J.E. 1985, MNRAS, 213, 435
\bibitem[\protect\citeauthoryear{Lacy}{2002}]{Lacy02} Lacy
M., 2002, ASPC, 290, 343
\bibitem[\protect\citeauthoryear{Lacy et al.}{2001}]{Lacy01} Lacy M.,
Laurent-Muehleisen S.A., Ridgway S.E., Becker R.H., White R.L., 2001, ApJ, 551, 17
\bibitem[\protect\citeauthoryear{Lacy et al.}{1999}]{Lacy99} Lacy M.,
Ridgway S.E., Wold M., Lilje P.B., Rawlings S., 1999, MNRAS, 307, 420
\bibitem[\protect\citeauthoryear{Lagos, Cora \& Padilla}{2008}]{Lagos08} Lagos
C.P., Cora S.A., Padilla N.D., 2008, MNRAS, 388, 587 
\bibitem[\protect\citeauthoryear{Laor}{2000}]{Laor00} Laor A., 2000, ApJ, 543, 111
\bibitem[\protect\citeauthoryear{Lindner \& Fragile}{2007}]{Lindner07} Lindner C.C., Fragile
P.C., 2007, AAS, 211, 4809
\bibitem[\protect\citeauthoryear{Lodato \& Pringle}{2006}]{Lodato06} Lodato G., Pringle J.E., 2006, MNRAS, 368, 1196
\bibitem[\protect\citeauthoryear{Magorrian et al.}{1998}]{Magorrian98} Magorrian J., Tremaine S., Richstone D., Bender R., Bower G., Dressler A.,
Faber S.M., Gebhardt K., et al., 1998, AJ, 115, 2285
\bibitem[\protect\citeauthoryear{Marulli et al.}{2008}]{Marulli08}
Marulli F., Bonomi S., Branchini E., Moscardini L., Springel V., 2008, MNRAS,
385, 1846
\bibitem[\protect\citeauthoryear{Merloni, Heinz \& di Matteo}{2003}]{Merloni03} Merloni A., Heinz S. di Matteo T., 2003, MNRAS, 345, 1057
\bibitem[\protect\citeauthoryear{Merloni \& Heinz}{2008}]{Merloni08} Merloni A., Heinz S., 2008, MNRAS, 388, 1011
\bibitem[\protect\citeauthoryear{Metcalf \& Magliocchetti}{2006}]{Metcalf06} Metcalf R.B., Magliocchetti M., 2006, MNRAS, 365, 101
\bibitem[\protect\citeauthoryear{Miller et al.}{1990}]{Miller90}
Miller L., Peacock J. A., Mead A.R.G., 1990, MNRAS, 244, 207
\bibitem[\protect\citeauthoryear{Moderski, Sikora \& Lasota}{1998}]{Moderski98} Moderski R., Sikora M., Lasota J.-P. 1998, MNRAS, 301, 142
\bibitem[\protect\citeauthoryear{Moderski \& Sikora}{1996}]{Moderski96} Moderski R., Sikora M. 1996, A\&AS, 120, 591
\bibitem[\protect\citeauthoryear{Nemmen et al.}{2007}]{Nemmen07} Nemmen R.S.,
Bower R.G., Babul A., Storchi-Bergmann T., 2007, MNRAS, 377, 1652
\bibitem[\protect\citeauthoryear{Nipoti, Blundell \& Binney}{2005}]{Nipoti05}
Nipoti C., Blundell M., Binney J., 2005, MNRAS, 361, 633
\bibitem[\protect\citeauthoryear{Omukai, Schneider \& Haiman}{2008}]{Omukai08}
Omukai K., Schneider R., Haiman Z., 2008, ApJ, 686, 801
\bibitem[\protect\citeauthoryear{Papaloizou \& Pringle}{1983}]{Papaloizou83}
Papaloizou  J.C.B., Pringle J.E., 1983, MNRAS, 202, 1181
\bibitem[\protect\citeauthoryear{Paz, Stasyszyn \& Padilla}{2008}]{Paz08} Paz
D, Stasyszyn F., Padilla N.D., 2008, MNRAS, 389, 1127
\bibitem[\protect\citeauthoryear{Peirani \& de Freitas}{2008}]{Peirani08} Peirani S., de Freitas Pacheco J.A., 2008, PhRvD, 77, 4023
\bibitem[\protect\citeauthoryear{Rezzolla et al.}{2008}]{Rezzolla08} Rezzolla
L., Barausse E., Dorband E.N., Pollney D., Reisswig C., Seiler J., Husa S.,
2008, PhRvD, 78d4002R
\bibitem[\protect\citeauthoryear{Shakura \& Sunyaev}{1973}]{Shakura73} Shakura
N.I., Sunyaev R.A., 1973, A\&A, 24, 337
\bibitem[\protect\citeauthoryear{Shapiro}{2005}]{Shapiro04} Shapiro S.L., 2005, ApJ, 620, 59
\bibitem[\protect\citeauthoryear{Sijacki et al.}{2007}]{Sijacki07} Sijacki D.,
Springel V., di Matteo T., Hernquist L., 2007, MNRAS, 380, 877
\bibitem[\protect\citeauthoryear{Sikora et al.}{2007}]{Sikora07} Sikora M.,
Stawarz L., Lasota J.-P., 2007, ApJ, 658, 815
\bibitem[\protect\citeauthoryear{Somerville et al.}{2008}]{Somerville08}
Somerville R. S., Hopkins P. F., Cox T. J., Robertson B. E., Hernquist L., 2008, arXiv:0808.1227
\bibitem[\protect\citeauthoryear{Spergel et al.}{2003}]{spergel03} Spergel D., et al.
(WMAP team),
2003, ApJS, 148, 175
\bibitem[\protect\citeauthoryear{Springel}{2005}]{Springel05}
Springel V., 2005, MNRAS, 364, 1105
\bibitem[\protect\citeauthoryear{Springel et al.}{2001}]{Springel01} Springel
V., White S., Tormen G., Kauffmann G., 2001, MNRAS, 328, 726
\bibitem[\protect\citeauthoryear{Spruit}{2008}]{spruit} Spruit H.C., 2008, arXiv:0804.3096
\bibitem[\protect\citeauthoryear{Sulentic et al.}{2003}]{Sulentic03} 
Sulentic J.W., Zamfir S., Marziani P., Bachev R., Calvani M.,
Dultzin-Hacyan D., 2003, ApJ, 597, L17
\bibitem[\protect\citeauthoryear{Ulvestad \& Ho}{2001}]{Ulvestad01} Ulvestad
J.S., Ho L.C., 2001, ApJ, 562, 133
\bibitem[\protect\citeauthoryear{Volonteri, Sikora \&
Lasota}{2007}]{Volonteri07} Volonteri M., Sikora M., Lasota J.-P., 2007, ApJ, 667, 704
\bibitem[\protect\citeauthoryear{Volonteri et al.}{2005}]{Volonteri05} Volonteri M., Madau P., Quataert E., Rees M.J., 2005, ApJ,
   620, 69
\bibitem[\protect\citeauthoryear{White et al.}{2007}]{White07} White R.L.,
Helfand D.J., Becker R.H., Glikman E., de Vries W., 2007, ApJ, 654, 99
\bibitem[\protect\citeauthoryear{Wilson \& Colbert}{1995}]{Further95} Wilson A.S., Colbert E.J.M., 1995, AAS, 186, 1502
\bibitem[\protect\citeauthoryear{Xu, Livio \& Baum}{1999}]{Xu99} Xu C., Livio
M., Baum S., 1999, AJ, 118, 1169
\bibitem[\protect\citeauthoryear{Wolf et al.}{2003}]{Wolf03} 
Wolf C., Wisotzki L., Borch A., Dye S., Kleinheinrich M., Meisenheimer K.,
2003, A\&A, 408, 499
\bibitem[\protect\citeauthoryear{Zamfir, Sulentic \& Marziani}{2008}]{Zamfir08}
Zamfir S., Sulentic J.W., Marziani P., 2008, MNRAS, 387, 856




\end{thebibliography}
\end{document}